\documentclass[
 amsmath,amssymb,
preprint,%
nofootinbib
]{revtex4}

\usepackage{bm}
\usepackage{makecell}
\usepackage{multirow}
\usepackage{graphicx}
\usepackage{epstopdf}
\usepackage{tablefootnote}
\usepackage{subfigure}
\usepackage{amsmath}
\usepackage{caption}
\usepackage{color}
\usepackage{enumerate}

\usepackage{lscape}

\renewcommand{\labelenumi}{\alph{enumi}:}

\newcommand{\la}{\langle}
\newcommand{\ra}{\rangle}
\newcommand{\B}{{\cal B}}
\newcommand{\A}{{\cal A}}
\newcommand{\ov}{\overline }
\newcommand{\Bc}{(\mathcal{B}_c)}
\newcommand{\Boct}{(\ov{\mathcal{B}}_8)}

\newcommand{\CP}{{\it CP}~}
\newcommand{\ACP}{\A_{C\!P}}
\newcommand{\CPV}{{\it CP\!V}~}
\newcommand{\be}{\begin{eqnarray}}
\newcommand{\en}{\end{eqnarray}}
\newcommand{\non}{\nonumber}
\newcommand{\lrpartial}{\buildrel\leftrightarrow\over\partial}

\begin{document}

\title{\CP Violation in Hadronic Weak Decays of Charmed Baryons\\ in the Topological Diagrammatic Approach}

\author{Hai-Yang Cheng}
\affiliation{Institute of Physics, Academia Sinica, Taipei, Taiwan 11529, Republic of China}

\author{Fanrong Xu}
\email{fanrongxu@jnu.edu.cn}
\author{Huiling Zhong}
 \affiliation{Department of Physics, College of Physics $\&$ Optoelectronic Engineering, Jinan University, Guangzhou 510632, P.R. China}


\small

\vskip 0.5cm
\begin{abstract}
\small
\vskip 0.5cm

\CP violation in the charmed baryon sector is generally expected to be very small, of order $10^{-4}$ or even smaller. Nevertheless,  large long-distance penguin topologies can be induced through final-state rescattering. 
At the hadron level, they are manifested as the triangle and bubble diagrams. However, these diagrams 
consist of not only the penguin but also other tree topologies. Therefore, we focus on the flavor 
structure of the these diagrams and project out their contributions to penguin and tree topologies. The analysis is performed within the framework of  the topological diagram approach and the irreducible SU(3) approach in which the SU(3) flavor symmetry of QCD is realized to describe the nonleptonic decays of charmed baryons.
\CP violation at the per mille level is found in the following decay modes:
$\Lambda_c^+ \to p \pi^0,~
\Lambda_c^+ \to p \eta', ~
\Xi_c^0 \to \Sigma^0\eta$ and 
$\Xi_c^+ \to \Sigma^+ \eta$.
\CP asymmetries in the last three modes receive sizable contributions from the SU(3)-flavor-singlet hairpin diagram.

\end{abstract}

\maketitle

\section{Introduction}
\label{intro}
The experimental and theoretical progresses in the study of hadronic decays of charmed baryons are truly very impressive in the past few years (for a review, see 
Ref. ~\cite{Cheng:2021qpd}). On the experimental side, around 44 measurements of  branching fractions and Lee-Yang parameters have been accumulated to date. On the theory aspect,  many dynamical model approaches had been developed in 1990s such as the relativistic quark model, the pole model and current algebra (see e.g.  \cite{Cheng:2021qpd}). However,  we still do not have a good phenomenological model, not mentioning the QCD-inspired approach which has been applied successfully to heavy meson decays, to describe the complicated physics of baryon decays. In the past decade, a very promising approach is to use the approximate SU(3) flavor symmetry of QCD to describe the two-body nonleptonic decays of charmed baryons.  There exist two distinct ways in realizing the flavor symmetry: the irreducible SU(3) approach (IRA) and the topological diagram approach (TDA). They provide a powerful tool for the model-independent analysis.  

In the IRA, SU(3) tensor invariants are constructed through the short-distance effective Hamiltonian, while in the TDA, the topological diagrams are classified according to the topologies in the flavor flow of weak decay diagrams with all strong-interaction effects included implicitly.   After 2014, the IRA became very popular
 \cite{Geng:2019xbo,Geng:2019awr,Geng:2020zgr,Huang:2021aqu,Zhong:2022exp,Xing:2023dni}.
The first analysis of two-body nonleptonic decays of antitriplet charmed baryons 
$\B_c(\bar 3)\to \B(8) P(8+1)$ within the framework of the TDA was performed by Kohara \cite{Kohara:1991ug}. A subsequent study was given by Chau, Cheng and Tseng in Ref. \cite{Chau:1995gk}  followed by some recent analyses in the TDA \cite{He:2018joe,Hsiao:2021nsc,Zhao:2018mov,Hsiao:2020iwc,Wang:2024ztg,Xing:2024nvg,Sun:2024mmk}. 
Although the TDA has been applied very successfully to charmed meson decays \cite{CC,Cheng:2016,Cheng:2024hdo}, its application to charmed baryon decays is more complicated than the IRA. As stressed in Ref. \cite{He:2018joe}, it is easy to determine the independent amplitudes in the IRA, while the TDA gives some redundancy. Some of the amplitudes are not independent and therefore should be absorbed into other amplitudes. Nevertheless, the TDA has the advantage that it is more intuitive, graphic and easier to implement model calculations.  The extracted topological amplitudes by fitting to the available data will enable us to probe the relative importance of different underlying decay mechanisms, and to relate one process to another at the topological amplitude level \cite{Zhong:2024qqs,Zhong:2024zme,Cheng:2024lsn}.

In this work we will focus on the hadronic weak decays of antitriplet charmed baryons into a octet baryon and a pseudoscalar meson: ${\cal B}_c(\bar 3)\to {\cal B}(8)P(8+1)$. Its general decay amplitude reads
\begin{eqnarray}
\label{eq:A&B}
M(\B_c\to \B_f+P)=i\bar u_f(A-B\gamma_5)u_c,
\end{eqnarray}
where $A$ and $B$ correspond to the parity-violating $S$-wave and parity-conserving $P$-wave amplitudes, respectively. Three Lee-Yang decay parameters $\alpha$, $\beta$ and $\gamma$ are of particular interest. They are defined by 
\begin{equation}
\begin{split}
& \alpha=\frac{2\kappa |A^*B|\cos(\delta_P-\delta_S)}{|A|^2+\kappa^2 |B|^2},~~
\beta=\frac{2\kappa |A^*B|\sin(\delta_P-\delta_S)}{|A|^2+\kappa^2 |B|^2},~~
\gamma=\frac{|A|^2-\kappa^2 |B|^2}{|A|^2+\kappa^2 |B|^2},
\end{split}
\label{eq:decayparameter}   
\end{equation}
with  $\kappa=p_c/(E_f+m_f)=\sqrt{(E_f-m_f)/(E_f+m_f)}$ and $\alpha^2+\beta^2+\gamma^2=1$. For completeness, the decay rate is given by
\begin{equation}
\label{eq:Gamma}
\Gamma = \frac{p_c}{8\pi}\frac{(m_i+m_f)^2-m_P^2}{m_i^2}\left(|A|^2
+ \kappa^2|B|^2\right). 
\end{equation}

There are two-fold purposes of the present work: (i) to update our previous analyses \cite{Zhong:2024qqs,Zhong:2024zme,Cheng:2024lsn} by incorporating the new data for global fits and most importantly
(ii) to explore the  {\it CP}-violating effects in the charmed baryon sector.  

As discussed in Ref.~\cite{Cheng:2024lsn}, the global fits to the data of $\Gamma$ and $\alpha$ usually encounter two major issues. First of all, there is a sign ambiguity for the decay parameter $\beta$ as $\alpha$ is proportional to $\cos(\delta_P-
\delta_S)$, while $\beta$ to $\sin(\delta_P-\delta_S)$. In other words, a measurement of $\alpha$ alone does not suffice to fix the phase shift between $S$- and $P$-wave amplitudes. Second, there is also a sign ambiguity for the decay parameter $\gamma$ as generally there exist two different sets of solutions for $A$ and $B$ based on the input of $\Gamma$.  It is obvious from Eqs. (\ref{eq:decayparameter}) and (\ref{eq:Gamma}) that one needs both information of $\Gamma$ and $\gamma$ to fix $|A|$ and $|B|$. 
Last year, LHCb has precisely measured the $\beta$ and $\gamma$ parameters of $\Lambda_c^+\to \Lambda \pi^+$ and $\Lambda_c^+\to \Lambda K^+$  decays for the first time \cite{LHCb:2024tnq}. These measurements will enable us to fix the phase shift $\delta_P-\delta_S$ and pick up the right solution for $S$- and $P$-wave amplitudes.   In principle, the accumulation of more data allows for the extraction of increasingly precise information.  After our latest analyisis
in Ref.~\cite{Cheng:2024lsn}, there are six  measurements of $\Xi_c^+$ decays
performed by Belle and Belle-II~\cite{Belle-II:2025klu,Belle:2024xcs}, which we will incorporate in this work
to perform more sensible global fits. 

The main purpose of this work is to study the  {\it CP}-violating effects 
in the charmed baryon sector {with SU(3) symmetry}.\footnote{
A recent discussion of possible SU(3)-breaking effects in hadronic charmed baryon decays can be found in Ref.~\cite{Yang:2025orn}.
}
Hence, we shall introduce various penguin contractions and write down the topological
penguin amplitudes in both TDA and IRA which transform as ${\bf 3}$ and are proportional to the CKM matrix element $\lambda_b=V_{cb}^*V_{ub}$. These amplitudes are denoted by  ${\cal A}^{\lambda_b}_{\rm TDA}$ and ${\cal A}^{\lambda_b}_{\rm IRA}$, respectively, below. In the TDA, 
the coefficients are related to penguin and tree topologies and hence they have clear graphic picture. The coefficients in ${\cal A}^{\lambda_b}_{\rm TDA}$ or ${\cal A}^{\lambda_b}_{\rm IRA}$ are unknown due to the lack of the experimental information on {\it CP}-violating observables. Nevertheless, they can be related to the 
tree topologies in ${\cal A}^{\rm tree}_{\rm TDA}$ or ${\cal A}^{\rm tree}_{\rm IRA}$
through final-state rescattering effects, a task which we will present in this work. Once  the coefficients are determined, we are allowed to investigate the rich {\it CP}-violating phenomenology in charmed baryon decays.  Moreover, \CP asymmetries are allowed to enhance 
from  $10^{-4}$ to the per mille level. 

The outline of this work is as follows. In Sec. II we introduce two different types of penguin contractions and write down the topological penguin amplitudes in both TDA and IRA. Sec. III is devoted to addressing the main topic of this work, namely, \CP violation in the charmed baryon sector. We begin with the first observation of the \CP asymmetry $\Delta \ACP$ in the charmed meson system reported by LHCb in 2019 and point out that successful 
numerical results and their implications are discussed in Sec. IV. Sec. V comes to our conclusions. Strong couplings of the pseudoscalar meson with baryons are given in Appendix A, 
currently available experimental data are collected in  Appendix B, while the new fitting
results of branching fractions, Lee-Yang parameters, magnitudes of $S$- and $P$-wave amplitudes and their phase shifts in the TDA  and IRA are presented in Appendix C.

\section{TDA and IRA}
\label{sec-1}

\begin{figure}[tp!]
\centering
\includegraphics[scale=0.58]{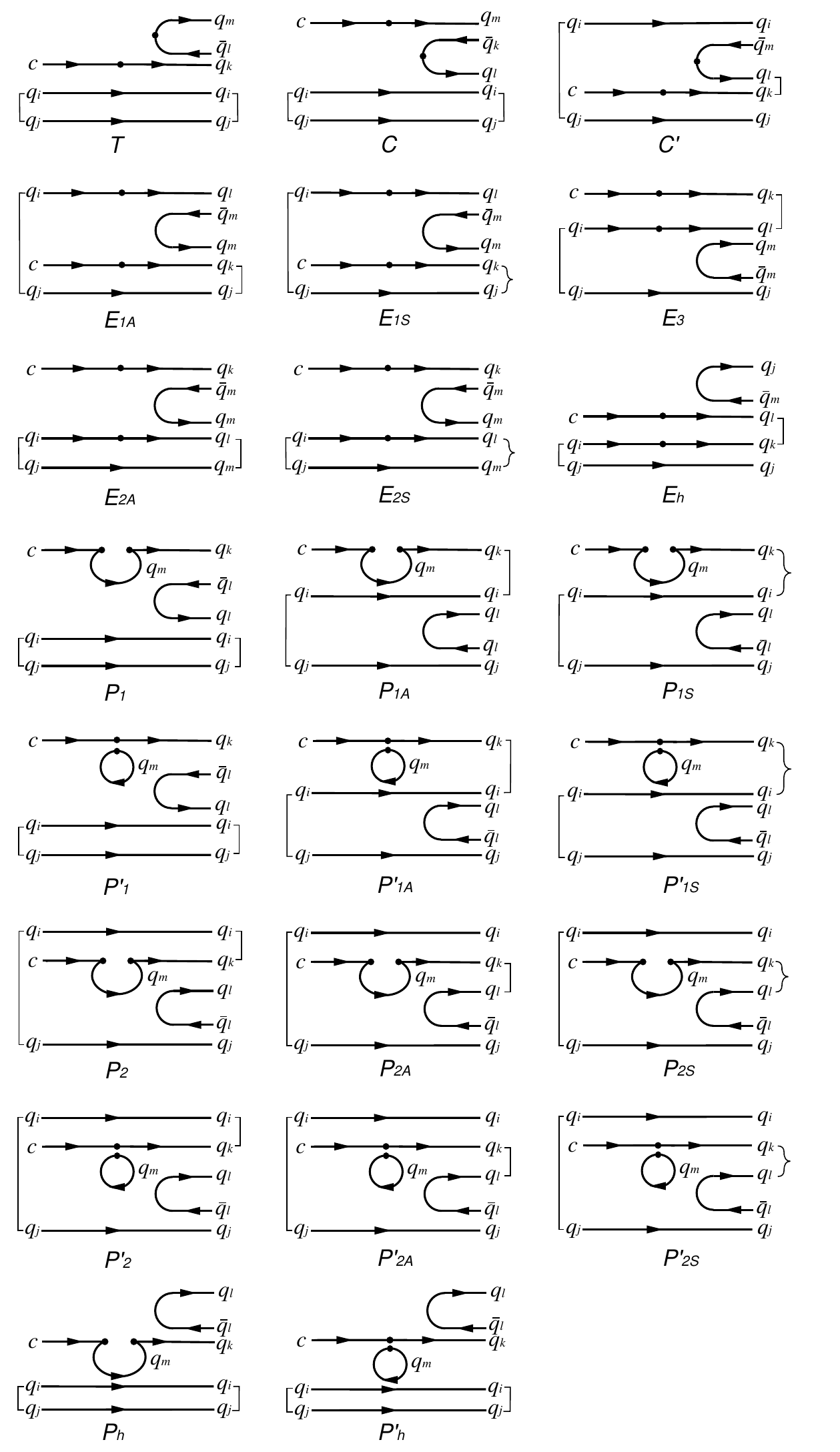}
\caption{Topological diagrams contributing to ${\cal B}_c(\bar 3)\to {\cal B}(8)P(8+1)$ decays. The symbols $]$ and $\}$ denote antisymmetric and symmetric two quark states, respectively. 
}
\label{Fig:TopDiag}
\end{figure}

There exist two distinct ways in realizing the approximate SU(3) flavor symmetry of QCD to describe the two-body nonleptonic decays of charmed baryons: the irreducible SU(3) approach (IRA) and the topological diagram approach (TDA). 

\subsection{TDA}
In terms of the octet baryon wave functions given by
\begin{eqnarray} \label{eq:wf8}
|{\cal B}^{m,k}(8)\rangle=a \,|\chi^m(1/2)_{A_{12}}\rangle|\psi^k(8)_{A_{12}}\rangle+ b\, |\chi^m(1/2)_{S_{12}}\rangle|\psi^k(8)_{S_{12}}\rangle
\end{eqnarray}
with $|a|^2+|b|^2=1$,  where the bases $\psi^k(8)_{A_{12}}$ and $\psi^k(8)_{S_{12}}$ for octet baryons denote the octet baryon states that are antisymmetric and symmetric in the first two quarks, respectively, and
$\chi^m(1/2)_{A,S}$ are the spin parts of the wave function defined in Eq. (23) of Ref. \cite{Chau:1995gk},
the relevant topological diagrams for the decays  of antitriplet charmed baryons ${\cal B}_c(\bar 3)\to {\cal B}(8)P(8+1)$ are depicted in Fig. \ref{Fig:TopDiag}: the external $W$-emission, $T$; the internal $W$-emission $C$; the inner $W$-emission $C'$; $W$-exchange diagrams $E_{1A}$, $E_{1S}$, $E_{2A}$, $E_{2S}$, 
$E_3$ and the hairpin diagram $E_h$. Since there are two possible penguin contractions, we will have penguin diagrams $P_{1(2)}$, $P_{1(2)A}$, $P_{1(2)S}$ as well as $P'_{1(2)}$, $P'_{1(2)A}$, $P'_{1(2)S}$. The topologies $P_h$ and $P'_h$ are hairpin penguin diagrams. 
The decay amplitudes of ${\cal B}_c(\bar 3)\to {\cal B}(8)P(8+1)$ in the TDA have the expressions \cite{Zhong:2024qqs,Zhong:2024zme}: \begin{equation}
\label{Eq:TDAamp}
\begin{aligned}
\mathcal{A}_{\rm T D A}
=
& \quad T ({\mathcal{B}}_c)^{i j} H_l^{k m}\left(\ov {\mathcal{B}}_8\right)_{i j k} (P^\dagger)_m^l\\
& +C (\mathcal{B}_c)^{i j} H_k^{m l}\left(\ov {\mathcal{B}}_8\right)_{i j l} (P^\dagger)_m^k
 + C' (\mathcal{B}_c)^{i j} H_m^{k l}\left(\ov {\mathcal{B}}_8\right)_{klj} (P^\dagger)_i^m \\
& +E_{1A} (\mathcal{B}_c)^{i j} H_i^{k l}\left(\ov {\mathcal{B}}_8\right)_{jkm} (P^\dagger)_l^m 
 + E_{1S} (\mathcal{B}_c)^{i j} H_i^{k l}(P^\dagger)_l^m \left[\left(\ov {\mathcal{B}}_8\right)_{jmk} 
+\left(\ov {\mathcal{B}}_8\right)_{kmj} \right] \\
& +E_{2A} (\mathcal{B}_c)^{i j} H_i^{k l}\left(\ov {\mathcal{B}}_8\right)_{jlm} (P^\dagger)_k^m   + E_{2S} (\mathcal{B}_c)^{i j} H_i^{k l} (P^\dagger)_k^m\left[\left(\ov {\mathcal{B}}_8\right)_{jml}
+ \left(\ov {\mathcal{B}}_8\right)_{lmj} \right] \\
&  +E_{3} (\mathcal{B}_c)^{i j} H_i^{k l}\left(\ov {\mathcal{B}}_8\right)_{klm} (P^\dagger)_j^m 
  +E_{h} (\mathcal{B}_c)^{i j} H_i^{k l}\left(\ov {\mathcal{B}}_8\right)_{klj} (P^\dagger)_m^m \\
&  +P_h (\mathcal{B}_c)^{i j} H_m^{m k}\left(\ov {\mathcal{B}}_8\right)_{ijk} (P^\dagger)_l^l 
  +P_{1} (\mathcal{B}_c)^{i j} H_m^{m k}\left(\ov {\mathcal{B}}_8\right)_{ijl} (P^\dagger)_k^l \\  
&  +P_{2A} (\mathcal{B}_c)^{i j} H_m^{m k}\left(\ov {\mathcal{B}}_8\right)_{kil} (P^\dagger)_j^l 
  +P_{2S} (\mathcal{B}_c)^{i j} H_m^{m k}(P^\dagger)_j^l \left[\left(\ov {\mathcal{B}}_8\right)_{kli}+
  \left(\ov {\mathcal{B}}_8\right)_{ilk}\right] \\ 
&  +P'_h (\mathcal{B}_c)^{i j} H_m^{k m}\left(\ov {\mathcal{B}}_8\right)_{ijk} (P^\dagger)_l^l 
  +P'_{1} (\mathcal{B}_c)^{i j} H_m^{k m}\left(\ov {\mathcal{B}}_8\right)_{ijl} (P^\dagger)_k^l \\  
&  +P'_{2A} (\mathcal{B}_c)^{i j} H_m^{k m}\left(\ov {\mathcal{B}}_8\right)_{kil} (P^\dagger)_j^l 
  +P'_{2S} (\mathcal{B}_c)^{i j} H_m^{k m}(P^\dagger)_j^l \left[\left(\ov {\mathcal{B}}_8\right)_{kli}+
  \left(\ov {\mathcal{B}}_8\right)_{ilk}\right],   
\end{aligned}
\end{equation}
where $(\mathcal{B}_c)^{ij}$ is an antisymmetric baryon matrix standing for antitriplet charmed baryons and  $P^{i}_{j}$ represent nonet mesons, 
\begin{equation}
\label{eq:B&P}
(\mathcal{B}_c)^{ij}=\left(\begin{array}{ccc}
 0 & \Lambda_c^+ & \Xi_c^+ \\
-\Lambda_c^+& 0 & \Xi_c^0 \\
-\Xi_c^+& -\Xi_c^0 & 0 
\end{array}\right), \qquad  P^{i}_{j}=\left(\begin{array}{ccc}
\frac{\pi^0 }{\sqrt{2}}+\frac{\eta_8}{\sqrt{6}}+\frac{\eta_1}{\sqrt{3}} & \pi^+ & K^+ \\
\pi^- & -\frac{\pi^0 }{\sqrt{2}}+\frac{\eta_8}{\sqrt{6}}+\frac{\eta_1}{\sqrt{3}} & K^0 \\
K^- & \overline{K}^0 & -\sqrt{2\over 3}{\eta_8}+\frac{\eta_1}{\sqrt{3}}\end{array}\right),
\end{equation}
with 
\begin{equation}
\label{eq:eta81qs}
    \eta_{8}=\sqrt{\frac{1}{3}}\eta_{q}-\sqrt{\frac{2}{3}}\eta_{s},\qquad
    \eta_{1}=\sqrt{\frac{2}{3}}\eta_{q}+\sqrt{\frac{1}{3}}\eta_{s}.
\end{equation}
The physical states $\eta$ and $\eta'$ are given by
\begin{equation}
\left(\begin{array}{c}
	\eta \\
	\eta'
\end{array}\right)=\left(\begin{array}{cc}
	\cos \phi & -\sin \phi\\
	\sin \phi & \cos \phi
\end{array}\right)\left(\begin{array}{l}
	\eta_{q} \\
	\eta_{s}
\end{array}\right)
=\left(\begin{array}{cc}
\cos \theta & -\sin \theta \\
\sin \theta & \cos \theta
\end{array}\right)\left(\begin{array}{c}
\eta_8 \\
\eta_1
\end{array}\right),
\end{equation}
where the mixing angles $\theta$ and $\phi$ are related through the relation $\theta=\phi-\arctan^{-1}\sqrt{2}$. 

The quark content of the octet baryons $\B(8)$ can be read from the subscript $ijk$ of the baryon tensor matrix
$(\mathcal{B}_8)_{i j k} = \epsilon_{ijl} (\mathcal{B}_8)^{l}_{k}$ with
\begin{equation}
(\mathcal{B}_8)^i_j=\left(\begin{array}{ccc}
\frac{1}{\sqrt{6}}\Lambda + \frac{1}{\sqrt{2}}\Sigma^0 & \Sigma^+ & p \\
\Sigma^- & \frac{1}{\sqrt{6}}\Lambda - \frac{1}{\sqrt{2}}\Sigma^0 & n \\
\Xi^-& \Xi^0& -\sqrt{\frac23}\Lambda \end{array}\right), \\
\end{equation}
The tensor coefficient $H^{kl}_m$ related to the CKM matrix elements appears in the standard model Hamiltonian with $H^{kl}_m(\bar q_k c)(\bar q_l q^m)$. The contraction of the two indices of $H^{kl}_m$, namely, $H^{m l}_m$, is induced in the penguin diagrams $P_1, P_{2A}, P_{2S}$ and $P_h$, while $H^{k l}_l$ in the penguin diagrams $P'_1, P'_{2A}, P'_{2S}$ and $P'_h$. 

In the diagrams $T$ and $C$, the two spectator quarks $q_i$ and $q_j$ are antisymmetric in flavor. 
Notice that the final-state quarks $q_l$ and $q_k$  in the topological diagrams $C'$, $E_3$ and $E_h$  also must be antisymmetric in flavor owing to the K\"orner-Pati-Woo (KPW) theorem which states that the quark pair in a baryon produced by weak interactions is required to be antisymmetric in flavor in the SU(3) limit \cite{Korner:1970xq}. Likewise, for $E_{1A,1S}$ and $E_{2A,2S}$, the KPW theorem also leads to 
\begin{equation}
\label{eq:E12A}
    E_{2A}=-E_{1A}, \qquad E_{2S}=-E_{1S}.
\end{equation} 
As a result, the number of independent  topological tree diagrams depicted in Fig. \ref{Fig:TopDiag} and the TDA tree amplitudes in Eq. (\ref{Eq:TDAamp}) is 7. 

Working out Eq. (\ref{Eq:TDAamp}) for ${\cal B}_c(\bar 3)\to {\cal B}(8)P(8+1)$ decays,  the obtained TDA decay amplitudes are listed in Tables I and II of Ref. \cite{Zhong:2024qqs} where the penguin contraction 
amplitudes have been neglected.
Among the 7 TDA tree amplitudes given in Eq. (\ref{Eq:TDAamp}), there still exist 2 redundant degrees of freedom through the redefinitions \cite{Chau:1995gk}:
\begin{eqnarray}
\label{eq:tildeTDA}
&& \tilde T=T-E_{1S}, \quad \tilde C=C+E_{1S},\quad \tilde C'=C'-2E_{1S}, \nonumber\\   
&& \tilde E_1=E_{1A}+E_{1S}-E_3, \quad \tilde E_h=E_h+2E_{1S}. 
\end{eqnarray}
As a result, among the seven topological tree amplitudes $T$, $C$, $C'$,  $E_{1A}$,  $E_h$, $E_{1S}$ and $E_3$, the last two are redundant degrees of freedom and can be omitted through redefinitions. 
It is clear that the minimum set of the topological tree amplitudes in the TDA is 5. This is in agreement with the number of tensor invariants found in the IRA \cite{Geng:2023pkr}.

In terms of the the decomposition of ${\bf 3\otimes\bar 3\otimes3}={\bf 3_p\oplus 3_t\oplus \bar 6\oplus 15}$ for the weak Hamiltonian responsible for $\Delta C=1$ weak transition,
the coefficient tensor $H$ in Eq. (\ref{Eq:TDAamp}) has the expression \cite{Grossman:2012ry}
\begin{equation}
\label{eq:Hijk}
H_k^{ij}=\frac12\left[
(H_{15})_k^{ij} +(H_{\ov 6})_k^{ij} \right] + \delta^j_k\left(\frac32 (H_{3p})^i-\frac12 (H_{3t})^i\right)+
 \delta^i_k\left(\frac32 (H_{3t})^j-\frac12 (H_{3p})^j\right).
\end{equation} 
For later purpose, it is also convenient to write
\begin{equation}
\begin{aligned}
\label{eq:Hp,Hm}
(H_+)^{ij}_k &\equiv H^{ij}_k+H^{ji}_k=(H_{15})^{ij}_k+(H_{3+})^i\delta^j_k+(H_{3+})^j\delta^i_k, \\
(H_-)^{ij}_k &\equiv H^{ij}_k-H^{ji}_k= (H_{\ov 6})^{ij}_k+2(H_{3-})^i\delta^j_k-2(H_{3-})^j\delta^i_k, \\
\end{aligned}
\end{equation}
with $H_{3\pm}=H_{3t}\pm H_{3p}$. 

In terms of $(\mathcal{B}_8)^{i}_{j}$ and $(\mathcal{B}_c)_{i}$ defined by $\frac12\epsilon_{ijk} (\mathcal{B}_c)^{jk}=(\Xi_c^0,-\Xi_c^+,\Lambda_c^+)$, the 
TDA amplitudes given in Eq. (\ref{Eq:TDAamp}) will have the decomposition \cite{Cheng:2024lsn}
\begin{equation}
\mathcal{A}_{\rm TDA}=\mathcal{A}_{\rm TDA}^{\rm tree}+\mathcal{A}_{\rm TDA}^{\lambda_b}
\end{equation}
with
\begin{equation}
\begin{aligned}
\label{eq:TDAtree2}
\mathcal{A}_{\rm TDA}^{\rm tree}= & \quad (T+C)\Bc_i\left(H_{15}\right)_m^{j l}\Boct_j^i (P^\dagger)_l^m 
 - E_h\Bc_i \left(H_{\ov 6}\right)_l^{i j}\Boct_j^l (P^\dagger)_m^m \\
& + (T-C-C'-2E_{1S})\Bc_i\left(H_{\ov 6}\right)_m^{j l}\Boct_j^i (P^\dagger)_l^m 
 -C'\Bc_i \left(H_{\ov 6}\right)_m^{i j}\Boct_j^l (P^\dagger)_l^m \\
& +(E_{1A}-E_{1S}-E_3)\Bc_i \left(H_{\ov 6}\right)_j^{i l}\Boct_m^j (P^\dagger)_l^m +2 E_{1S}\Bc_i \left(H_{\ov 6}\right)_m^{j l}\Boct_j^m (P^\dagger)_l^i  \\
\end{aligned}
\end{equation}
and 
\begin{equation}
\begin{aligned}
\label{eq:TDAlambdab2}
\mathcal{A}_{\rm TDA}^{\lambda_b}=  & \quad \tilde b_1 \Bc_i\left(H_3\right)^j\Boct_j^i (P^\dagger)_l^l
+\tilde b_2 \Bc_i\left(H_3\right)^j\Boct_j^l (P^\dagger)_l^i
+\tilde b_3 \Bc_i\left(H_3\right)^i\Boct_j^l (P^\dagger)_l^j \\
& +\tilde b_4 \Bc_i\left(H_3\right)^l\Boct_j^i (P^\dagger)_l^j 
 +\tilde b_5 \Bc_i\left(H_{{15}^b}\right)^{jk}_l\Boct_j^i (P^\dagger)_k^l, \\
\end{aligned}
\end{equation}
where the matrix element $H_3$ is normalized to  $(1,0,0)\lambda_b$ and  the coefficients $\tilde b_i$ can be expressed in terms of tilde quantities \cite{Cheng:2024lsn}
\footnote{ The definitions of $\tilde{b}_i$ $(i=1,\ldots, 4)$ have been consistently refined relative to those in Ref.~\cite{Cheng:2024lsn}. }
\begin{equation}
\begin{aligned}
\label{eq:bprime}
& \tilde b_1=\frac14(\tilde T-3\tilde  C)-\frac12 \tilde C'-\frac12 \tilde E_h-2\tilde P_h-2\tilde P_{2S}, \\
& \tilde b_2= \frac12 \tilde C' +2\tilde P_{2S}, \\
& \tilde b_3= -\frac12 \tilde C'+\frac12\tilde E_{1A}-\tilde P_{2A}-\tilde P_{2S},\\
& \tilde b_4=  -\frac14(3 \tilde T-\tilde C)+\frac12\tilde C'+\frac12\tilde E_{1}-2\tilde P_1+\tilde P_{2A}+\tilde P_{2S}, \\
&  \tilde b_5=\tilde T+\tilde C,  \\
\end{aligned}
\end{equation} 
with 
\begin{equation}
\begin{aligned}
\tilde P_1 & = P_1+E_{1S}-\frac12 E_3, \qquad   \tilde P_h  = P_h - E_{1S}, \\
\tilde P_{2S} & = P_{2S}+E_{1S} , \qquad  \qquad ~\tilde P_{2A}  = P_{2A}+E_{1S}-E_3, \\
\tilde P'_1 & = P'_1-E_{1S}+\frac12 E_3, \qquad  \tilde P'_h  = P'_h + E_{1S}, \\
\tilde P'_{2S} & = P'_{2S}-E_{1S} , \qquad  \qquad ~\tilde P'_{2A}  = P'_{2A}-E_{1S}+E_3.  
\end{aligned}
\end{equation} 
In Eq. (\ref{eq:TDAlambdab2}) we have followed Ref. \cite{He:2024pxh} to lump the components of $(H_{15})^{ij}_k$  proportional to $\lambda_b$ into $(H_{{15}^b})^{ij}_k$.
The $\mathcal{A}_{\rm TDA}^{\lambda_b}$ involves penguin-like contributions induced from the current-current
tree operators $O_{1,2}$ or penguin contributions arising from QCD penguin operators.  In the TDA, the relations of $\tilde b_1,\cdots, \tilde b_4$ with the penguin diagrams depicted in Fig. \ref{Fig:TopDiag} 
are explicitly shown in Eq. (\ref{eq:bprime}).

From the expression of  $\mathcal{A}_{\rm TDA}^{\rm tree}$ in Eq. (\ref{eq:TDAtree2}), we see  that there is only one term related to $H_{15}$.  However, there are only five independent terms in $\mathcal{A}_{\rm TDA}^{\rm tree}$. 
This is because the last five terms are not totally independent. Indeed, $\mathcal{A}_{\rm TDA}^{\rm tree}$ is one of the general IRA amplitudes given in Eq. 
(\ref{eq:IRAamp}) below.  We shall show that the number of the minimum set of tensor invariants in $\mathcal{A}_{\rm TDA}^{\rm tree}$ is 5.

\subsection{IRA}
The general SU(3) invariant decay amplitudes in the IRA read \cite{He:2018joe}
\begin{equation}
\begin{aligned}
\label{eq:IRAamp}
\mathcal{A}_{\rm IRAa}= & \quad a_{1}\left({\B}_c\right)_i\left(H_{\ov 6}\right)_j^{i k}\left(\ov{\B}_8\right)_k^j (P^\dagger)_l^l
+a_{2}\left({\B}_c\right)_i\left(H_{\ov 6}\right)_j^{i k}\left(\ov{\B}_8\right)_k^l (P^\dagger)_l^j
+a_{3}\left({\B}_c\right)_i\left(H_{\ov 6}\right)_j^{i k}\left(\ov{\B}_8\right)_l^j (P^\dagger)_k^l \\
& +a_{4}\left({\B}_c\right)_i\left(H_{\ov 6}\right)_l^{j k}\left(\ov{\B}_8\right)_j^i (P^\dagger)_k^l
+a_{5}\left({\B}_c\right)_i\left(H_{\ov 6}\right)_l^{j k}\left(\ov{\B}_8\right)_j^l (P^\dagger)_k^i
+a_{6}\left({\B}_c\right)_i\left(H_{15}\right)_j^{i k}\left(\ov{\B}_8\right)_k^j (P^\dagger)_l^l \\
& +a_{7}\left({\B}_c\right)_i\left(H_{15}\right)_j^{i k}\left(\ov{\B}_8\right)_k^l (P^\dagger)_l^j 
+a_{8}\left({\B}_c\right)_i\left(H_{15}\right)_j^{i k}\left(\ov{\B}_8\right)_l^j (P^\dagger)_k^l
+a_{9}\left({\B}_c\right)_i\left(H_{15}\right)_l^{j k}\left(\ov{\B}_8\right)_j^i (P^\dagger)_k^l\\
& +a_{10}\left({\B}_c\right)_i\left(H_{15}\right)_l^{j k}\left(\ov{\B}_8\right)_j^l (P^\dagger)_k^i
+b_1 \Bc_i\left(H_3\right)^j\Boct_j^i (P^\dagger)_l^l
+b_2 \Bc_i\left(H_3\right)^j\Boct_j^l (P^\dagger)_l^i \\
& +b_3 \Bc_i\left(H_3\right)^i\Boct_j^l (P^\dagger)_l^j 
 +b_4 \Bc_i\left(H_3\right)^l\Boct_j^i (P^\dagger)_l^j,\\
\end{aligned}
\end{equation}
where the last four terms related to penguins are taken from Ref. \cite{He:2024pxh}. 
The first five terms associated with $H_{\ov 6}$ are not totally independent as one of them is redundant through the redefinition.  For example, the following two redefinitions 
\begin{equation}
\label{eq:redef2}
 a_1'=a_1-a_5, \quad   a_2'=a_2+a_5, \quad a_3'=a_3+a_5, \quad a_4'=a_4+a_5,  
\end{equation} 
and \cite{He:2018joe}
\begin{equation}
\label{eq:redef1}
 a_1''=a_1+a_2, \quad   a_2''=a_2-a_3, \quad a_3''=a_3-a_4, \quad a_5''=a_5+a_3.  
\end{equation}
are possible. As for the five terms associated with $H_{\overline{15}}$ in Eq. (\ref{eq:IRAamp}), four of them are prohibited by the KPW theorem together with the pole model, namely, $a_6=a_7=a_8=a_{10}=0$ \cite{Geng:2018rse,Geng:2023pkr}.

Comparing the expression of  $\mathcal{A}_{\rm TDA}^{\rm tree}$ given in Eq. (\ref{eq:TDAtree}) with
$\mathcal{A}_{\rm IRAa}$ we have
\begin{equation}
\begin{aligned}
& a_1=- E_h, \qquad a_2=-C', \qquad a_3= E_{1A}-E_{1S}-E_3, \\
& a_4= T-C-C'-2E_{1S}, \qquad a_5=2E_{1S}, \qquad a_9=T+C. 
\end{aligned}
\end{equation}
The redefinition of Eq. (\ref{eq:redef2}) yields
\begin{equation}
\begin{aligned}
& a_1'=- \tilde{E}_h, \qquad a_2'= -\tilde C', \qquad a_3'= \tilde{E}_1,\\
& a_4'= \tilde T-\tilde C-\tilde C',
\qquad a_9=\tilde T+\tilde C, \\
\end{aligned}
\end{equation}
with the tilde parameters defined in Eq. (\ref{eq:tildeTDA}). Hence, the minimum set of the topological tree amplitudes in the TDA is 5, namely, $\tilde T$, $\tilde C$, $\tilde C'$, $\tilde{E}_1$ and $\tilde{E}_h$. 

There is another set of the IRA amplitudes given in Refs. \cite{Geng:2023pkr} and \cite{He:2024pxh}:
\be
\mathcal{A}_{\rm IRAb}=\mathcal{A}_{\rm IRAb}^{\rm tree}+\mathcal{A}_{\rm IRAb}^{\lambda_b}
\en
with
\be
\label{eq:IRAb}
\mathcal{A}_{\rm IRAb}^{\rm tree} &= &  \tilde{f}^a\left(\ov {\B}_c\right)^{ik}\left(H_{\ov 6}\right)_{ij}\left({\B}_8\right)_k^j (P^\dagger)_l^l+\tilde{f}^b \left({\B}_c\right)^{ik}\left(H_{\ov 6}\right)_{ij}\left(\ov{\B}_8\right)_k^l (P^\dagger)_l^j+\tilde{f}^c \left({\B}_c\right)^{ik}\left(H_{\ov 6}\right)_{ij}\left(\ov{\B}_8\right)_l^j 
(P^\dagger)_k^l \non \\
&+&  \tilde{f}^d \left({\B}_c\right)^{kl}\left(H_{\ov 6}\right)_{ij}\left(\ov{\B}_8\right)_k^i M_l^j+\tilde{f}^e\left({\B}_c\right)_j\left(H_{15}\right)_l^{i k}\left(\ov{\B}_8\right)_i^j (P^\dagger)_k^l, 
\en
and
\begin{equation}
\begin{aligned}
\label{eq:IRAHe}
\mathcal{A}_{\rm IRAb}^{\lambda_b}=  & \quad \tilde{f}^a_3 \Bc_i\left(H_3\right)^j\Boct_j^i (P^\dagger)_k^k
+\tilde{f}^b_3 \Bc_k\left(H_3\right)^i\Boct_i^l (P^\dagger)_l^k 
 +\tilde{f}^c_3 \Bc_i\left(H_3\right)^i\Boct_j^l (P^\dagger)_l^j \\
&  +\tilde{f}^d_3 \Bc_i\left(H_3\right)^l\Boct_j^i (P^\dagger)_l^j +\tilde{f}^e \Bc_i\left(H_{{15}^b}\right)^{jk}_l\Boct_j^i (P^\dagger)_k^l.  \\
\end{aligned}
\end{equation}
The advantage of this IRAb is that both $\mathcal{A}_{\rm IRAb}^{\rm tree}$ and  $\mathcal{A}_{\rm IRAb}^{\lambda_b}$ have the minimum sets of tensor invariants, namely, 5 in the former and 4 in the latter. 
The equivalence between $\widetilde{\rm TDA}$, IRAa and IRAb leads to the relations:
\begin{equation}
\begin{aligned}
\label{eq:tildeTDAIRA1}
&\Tilde{T}
=\frac{1}{2}( \tilde f^b+\tilde f^e)=\frac12(-a'_2+a'_4+a_9),
\qquad
\Tilde{C}
=\frac{1}{2}(-\tilde f^b+\tilde f^e)=\frac12(a'_2-a'_4+a_9), \\
& \Tilde{C'}=\tilde f^b-\tilde f^d=-a'_2,\qquad
\Tilde{E_{1}}=-\tilde f^c=a'_3,\qquad
\Tilde{E_{h}}=\tilde f^a=-a'_1,\\
\end{aligned}
\end{equation}
or inversely
\begin{equation}
\begin{aligned}
\label{eq:tildeTDAIRA2}
& \tilde f^a=\tilde E_h=-a'_1, \qquad \tilde f^b=\tilde T-\tilde C=-a'_2+a'_4, \qquad \tilde f^c=-\tilde E_1=-a'_3, \\
& \tilde f^d=\tilde T-\tilde C-\tilde C'=-a'_4, \qquad \tilde f^e=\tilde T+\tilde C=a_9.
\end{aligned}
\end{equation}

\section{$CP$ violation}
According to the standard model, {\it CP} violation is at a very small level in the decays of charmed hadrons. This is mainly because of the relation of the CKM matrix elements, $\lambda_s\approx -\lambda_d$. As a consequence, {\it CP} violation in the charm sector is usually governed by $\lambda_b$ which is very tiny compared to $\lambda_d$ or $\lambda_s$ in magnitude. This also indicates that the corresponding QCD penguin and electroweak penguin are also rather suppressed.  In order to have direct {\it CP} asymmetries in partial rates, it is necessary to have nontrivial weak and strong phase differences.
\subsection{\CP violation in the charmed meson sector}

In November of 2011 LHCb announced the first evidence of \CP violation in the charm sector. A nonzero value for the difference between the time-integrated \CP asymmetries of the decays $D^0\to K^+K^-$ and $D^0\to\pi^+\pi^-$ \cite{LHCb:2011osy}
\begin{equation} \label{eq:LHCb:2011}
\Delta \ACP\equiv \A_{CP}(K^+K^-)-\A_{CP}(\pi^+\pi^-)=(-0.82\pm0.21\pm0.11)\%
\end{equation}
was reported. The significance of the measured deviation from zero is 3.5$\sigma$. This had triggered a flurry of studies exploring whether this \CP violation in the charm sector was consistent with the standard model (SM) or implies new physics (NP). However, the original evidence of \CP asymmetry difference was gone in 2013 and 2014 when LHCb started to use the muon tag to identify the $D^0$ and found a positive $\Delta \A_{C\!P}$ \cite{LHCb:2013dkm}.
In 2019, LHCb finally reported the first observation of
\CP asymmetry in the charm system with the result at the per mille level \cite{LHCb:CP}
\begin{equation} \label{eq:LHCb:2019}
\Delta \ACP=(-1.54\pm0.29)\times 10^{-3}.
\end{equation}

In the standard-model estimate with the short-distance penguin contribution, we have the expression (see e.g. \cite{Cheng:2012a})
\begin{eqnarray}
\Delta \ACP &=&  -{4{\rm Im}[(\lambda_s-\lambda_d)\lambda_b^*]\over |\lambda_s-\lambda_d|^2}  \left( \left|{P\over T+E}\right|_{_{K\!K}}\sin\theta_{_{K\!K}}+ \left|{P\over T+E}\right|_{_{\pi\pi}}\sin\theta_{_{\pi\pi}} \right), \\
&=& -1.31\times 10^{-3}  \left( \left|{P\over T+E}\right|_{_{K\!K}}\sin\theta_{_{K\!K}}+ \left|{P\over T+E}\right|_{_{\pi\pi}}\sin\theta_{_{\pi\pi}} \right), 
\end{eqnarray}
where $\theta{_{K\!K}}$ is the strong phase of $(P/T)_{_{K\!K}}$ and likewise for $\theta{_{\pi\pi}}$. Since 
$|P/T|$ is na{\"i}vely expected to be of order $(\alpha_s(\mu_c)/\pi)\sim {\cal O}(0.1)$, it appears that $\Delta \ACP$ is most likely of order $10^{-4}$ assuming strong phases close to $90^\circ$ or even less for realistic strong phases.  It was pointed out in Ref. \cite{Cheng:2012a} that there is a resonant-like final-state rescattering which has the same topology as the QCD-penguin. That is, the penguin topology receives sizable long-distance contributions through final-state interactions. In 2012 an ansatz that the long-distance penguin $P^{\rm LD}$ is of the same order of magnitude as $E$  was made in Ref. \cite{Cheng:2012a}.  Since the $W$-exchange topology can be extracted from the data, this implies that one can make a reliable prediction of $\Delta \ACP$. Based on  the ansatz of $P^{\rm LD}=E^{\rm LD}\approx E$, it was predicted in Ref. \cite{Cheng:2012b}
 in 2012 that $\Delta\ACP=(-0.151\pm0.004)\%$ and $(-0.139\pm0.004)\%$ for the two solutions of $W$-exchange amplitudes.  They are amazingly in excellent agreement with the LHCb observation of  \CP violation in the charm meson sector in 2019. 
\footnote{Based on the pQCD calculation of the short-distance penguin contributions, it was claimed in the so-called factorization-assisted topological amplitude (FAT) approach that $\Delta\ACP\approx -1.00\times 10^{-3}$~\cite{Li:2012cfa}. Although both FAT+pQCD and QCDF+TDA  approaches predict a \CP asymmetry difference at the per mille level, the mechanisms responsible for \CP violation are quite different. It comes from the interference between the tree $T+E$ and SD penguin $P$ in the former approach, while it is the interference between tree and LD penguin amplitudes that pushes $\Delta \ACP$ to the $10^{-3}$ level in the latter.
 As discussed in Refs. \cite{Cheng:2019ggx,Cheng:2021yrn}, both approaches can be easily discriminated in the $D\to V\!P$ sector.}
Hence,  it is the interference between tree and long-distance penguin amplitudes that pushes $\Delta \A_{CP}$ up to the $10^{-3}$ level \cite{Cheng:2012b}. We also see that the LHCb's observation of \CP violation can be explained within the SM without the need of NP.

\begin{figure}[t]
\centering
\includegraphics[scale=0.60]{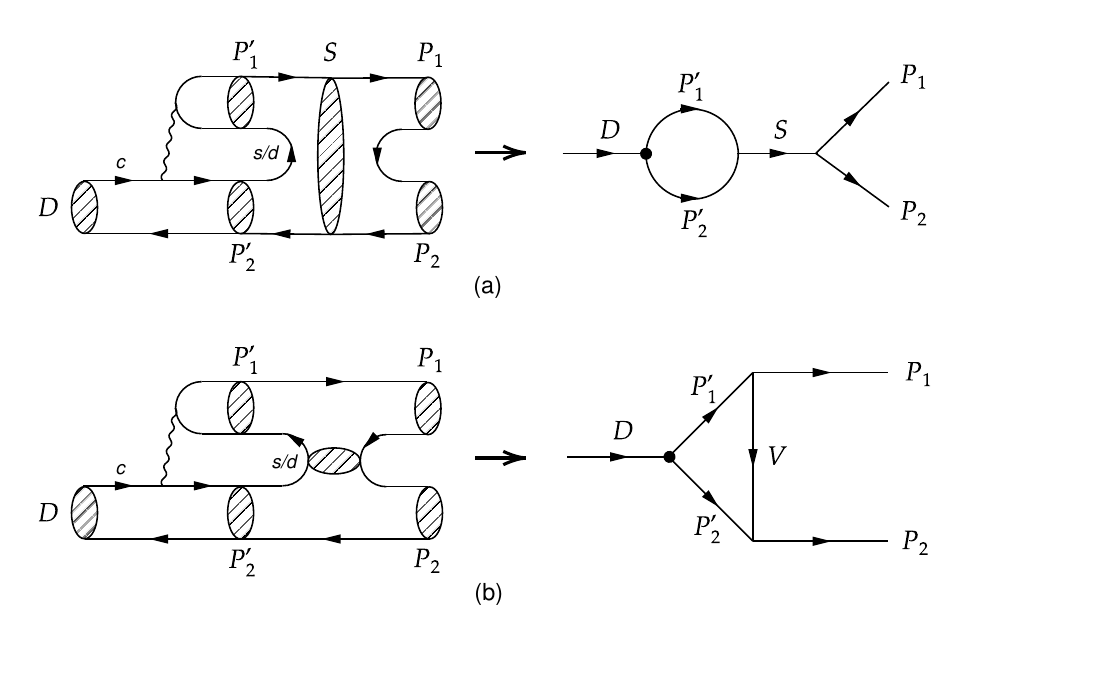}
\caption{\small Final-state $s$- and $t$-channel rescattering contributions to $D\to P_1P_2$ from the external $W$-emission.  Pole and triangle diagrams are the corresponding processes at the hadron level.
}
\label{Fig:FSI_meson}
\end{figure}

In the topological diagrammatic approach there exists two possible final-state rescattering processes of the short-distance $T$  which have the penguin topology in $s$- and $t$-channel diagrams (see Fig. \ref{Fig:FSI_meson}). It was shown
in Ref. \cite{Wang:2021rhd} that the ansatz of $P^{\rm LD}=E^{\rm LD}$ was justified through a  systematical
study of the triangle diagrams  corresponding to the $t$-channel rescattering. The $s$-channel rescattering can proceed through the intermediate scalar and vector resonances, but the existence of the nearby scalar resonances
close to the charmed meson mass such as $f_0(1710)$ and $f_0(1790)$ implies that the $s$-channel is mainly governed by scalar resonances. Likewise, 
the relation of $P^{\rm LD}=E^{\rm LD}$ in the $s$-channel rescattering also holds in the SU(3) limit as we are going to show below. 

Consider the $s$-channel process $D\to P'_1P'_2\to S\to P_1P_2$ in Fig. \ref{Fig:FSI_meson}(a). 
\begin{equation}
\begin{split}
\la S|H_{\rm eff}|D\ra=\sum_{P'_1,P'_2}D_i H^{jk}_l(P'^\dagger_1)^l_j (P'^\dagger_2)^i_k\left[ (P'_1)^m_n(P'_2)^n_o(S^\dagger)^o_m+
(P'_1)^m_n(P'_2)^o_m(S^\dagger)^n_o\right], \\
\end{split}
\end{equation} 
where $P'_1$ and $P'_2$ are summed over. Since $P={1\over \sqrt{2}}\sum_{a=1}^8 P_a\lambda_a$, we then have
\begin{equation}
\sum_{P'_1}(P'^\dagger_1)^l_j(P'_1)^m_n=\frac12\sum_{\lambda_a}(\lambda_a^\dagger)^l_j (\lambda_a)^m_n=\delta^l_n\delta^m_j-\frac13 \delta^l_j\delta^m_n.
\end{equation}
Hence,
\begin{equation}
\begin{split}
\la S|H_{\rm eff}|D\ra 
& =D_i\left[ H^{jl}_l(S^\dagger)^i_j -\frac13 H^{ji}_o(S^\dagger)^o_j-\frac13 H^{lm}_l(S^\dagger)^i_m
+\frac19 H^{li}_l(S^\dagger)^m_m \right] \\
& + D_i\left[ H^{ik}_l(S^\dagger)^l_k -\frac13 H^{ji}_l(S^\dagger)^l_j-\frac13 H^{lk}_l(S^\dagger)^i_k
+\frac19 H^{li}_l(S^\dagger)^n_n \right]. \\
\end{split}
\end{equation} 
Note that the 4-quark operators can be written as
\begin{equation}
\begin{split}
c_1(\bar q_i c)(\bar q_j q^k) + c_2 (\bar q_j c)(\bar q_i q^k) & 
 =c_+[(\bar q_i c)(\bar q_j q^k)+(\bar q_j c)(\bar q_i q^k)] +c_-[(\bar q_i c)(\bar q_j q^k)-(\bar q_j c)(\bar q_i q^k)] \\
 & = c_+ (H_+)^{ij}_k(\bar q_i c)(\bar q_j q^k)+ c_-(H_-)^{ij}_k(\bar q_i c)(\bar q_j q^k),\\
\end{split}
\end{equation} 
where the SU(3) tensor $(H_+)^{ij}_k$ ($(H_-)^{ij}_k$) is symmetric (antisymmetric) in the $ij$ indices. 
It follows that
\begin{equation}
\la S|H_{\rm eff}|D \ra= \frac13 c_+\left [D_i(H_+)^{ik}_l (S^\dagger)^l_k+
D_i(H_+)^{jl}_l (S^\dagger)^i_j \right]+ \frac53 c_-\left [D_i(H_-)^{ik}_l (S^\dagger)^l_k+
D_i(H_-)^{jl}_l (S^\dagger)^i_j \right]
\end{equation}
The first term in [...] is a $W$-exchange and the second term a penguin with equal weight. Hence, the final-state rescattering of  SD $T$ will yield a topology of $W$-exchange and the penguin $P$. This implies that the $s$-channel diagram also yields the desired relation $P^{\rm LD}=E^{\rm LD}$.

As it has been realized that SD penguin contributions are expected to play a less important role towards the understanding of the measured \CP asymmetry difference, LD effects such as the final-state rescattering processes   $\pi\pi\to \pi\pi$, $\pi K\to \pi K$ and $\pi\pi\to K\ov K$ have been studied in detail in Refs.~\cite{Pich:2023kim,Bediaga:2022sxw}. However, the predicted level of \CP violation is much below the experimental value.  The above-mentioned rescattering effects turn out not producing enough enhancement.  
Likewise, calculations based on QCD light-cone sum rules \cite{Lenz:2023rlq} cannot account for the LHCb measurement. 
It appears that the TDA is so far the only approach in which the LD penguin contribution can be estimated in a reliable way. It is a data-driven approach which is suitable for tackling with the nonperturbative QCD effects in charm decays. Indeed, the TDA has been applied very successfully to charmed meson decays \cite{CC,Cheng:2016,Cheng:2024hdo}. 
In the TDA, the LD rescattering contribution to the penguin is of the same order as $W$-exchange $E$. Hence, the large penguin induced from the final-state rescattering is adequate to account for the \CP violation observed by LHCb.

\subsection{\CP violation in the charmed baryon sector}
\subsubsection{Observables} 
In the charmed baryon sector, besides {\it CP} asymmetries in partial decay rates, two other 
{\it CP}-violation quantities of interest are:
\begin{equation}
\A_\alpha={ \alpha+\bar\alpha \over \alpha-\bar\alpha}, \qquad R_\beta={ \beta+\bar\beta \over \alpha-\bar\alpha}.
\end{equation}
They are related to weak and strong phase differences between $S$- and $P$-wave amplitudes, $\Delta\phi$ and $\Delta\delta\equiv \delta_P-\delta_S$, respectively, via
\begin{equation}
\A_\alpha=-\tan\Delta\delta\tan\Delta\phi, \qquad R_\beta=\tan\Delta\phi.
\end{equation}
{\it CP} asymmetries $\A_\alpha$ and $R_\beta$ for $\Lambda_c^+\to\Lambda \pi^+, \Lambda K^+$ and $\Lambda_c^+\to pK_S^0$ have been measured recently by LHCb with null results \cite{LHCb:2024tnq}.

The existence of strong phases in the partial-wave amplitudes of hadronic charmed baryon decays plays a pivotal role in a further exploration of {\it CP} violation in the charmed baryon decays.   
We consider the singly-Cabibbo-suppressed charmed baryon decay with the amplitude $A=\lambda_d A_d+\lambda_s A_s$. Its {\it CP} asymmetries defined by
\begin{equation}
\ACP\equiv { \Gamma(\B_c\to \B P)-\Gamma(\ov{\B}_c\to \bar{\B}\bar P)\over 
\Gamma(\B_c\to \B P)+\Gamma(\ov{\B}_c\to \bar{\B}\bar P) }, \qquad \ACP^S={S-\bar S\over S+\bar S}, \qquad \ACP^P={P-\bar P\over P+\bar P},
\end{equation}
have the expression
\begin{equation}
\begin{aligned}
\ACP& =  {2 {\rm Im}(\lambda_d\lambda_s^*)\over |\lambda_d|^2} \,
{ {\rm Im}(A_dA_s^*)\over |A_d-A_s|^2}
 = 1.31\times 10^{-3} {|A_d A_s|\over |A_d-A_s|^2} \sin\delta_{ds}, \\
\end{aligned}
\end{equation}
for the partial rate asymmetry,
where $\delta_{ds}$ is the strong phase of $A_s$ relative to $A_d$. If the decay amplitude is written as
$A=\frac12(\lambda_s-\lambda_d)(A_s-A_d)-\frac12\lambda_b(A_s+A_d)$, we will have
\begin{equation}
\begin{aligned}
\A_{CP} & = - {2 {\rm Im}[(\lambda_s-\lambda_d)\lambda_b^*]\over |\lambda_s-\lambda_d|^2} \,
{ {\rm Im}[(A_s-A_d)(A_s^*+A_d^*)]\over |A_s-A_d|^2} \\
 & =  {4 {\rm Im}[(\lambda_s-\lambda_d)\lambda_b^*]\over |\lambda_s-\lambda_d|^2} { {\rm Im}(A_dA_s^*)\over |A_s-A_d|^2}= 1.31\times 10^{-3} {|A_d A_s|\over |A_s-A_d|^2} \sin\delta_{ds}, \\
\end{aligned}
\end{equation}
which is the same as before.
In general, {\it CP} asymmetry is expected to be of order $10^{-4}$ or even smaller. 

\begin{figure}[t]
\centering
\includegraphics[scale=0.60]{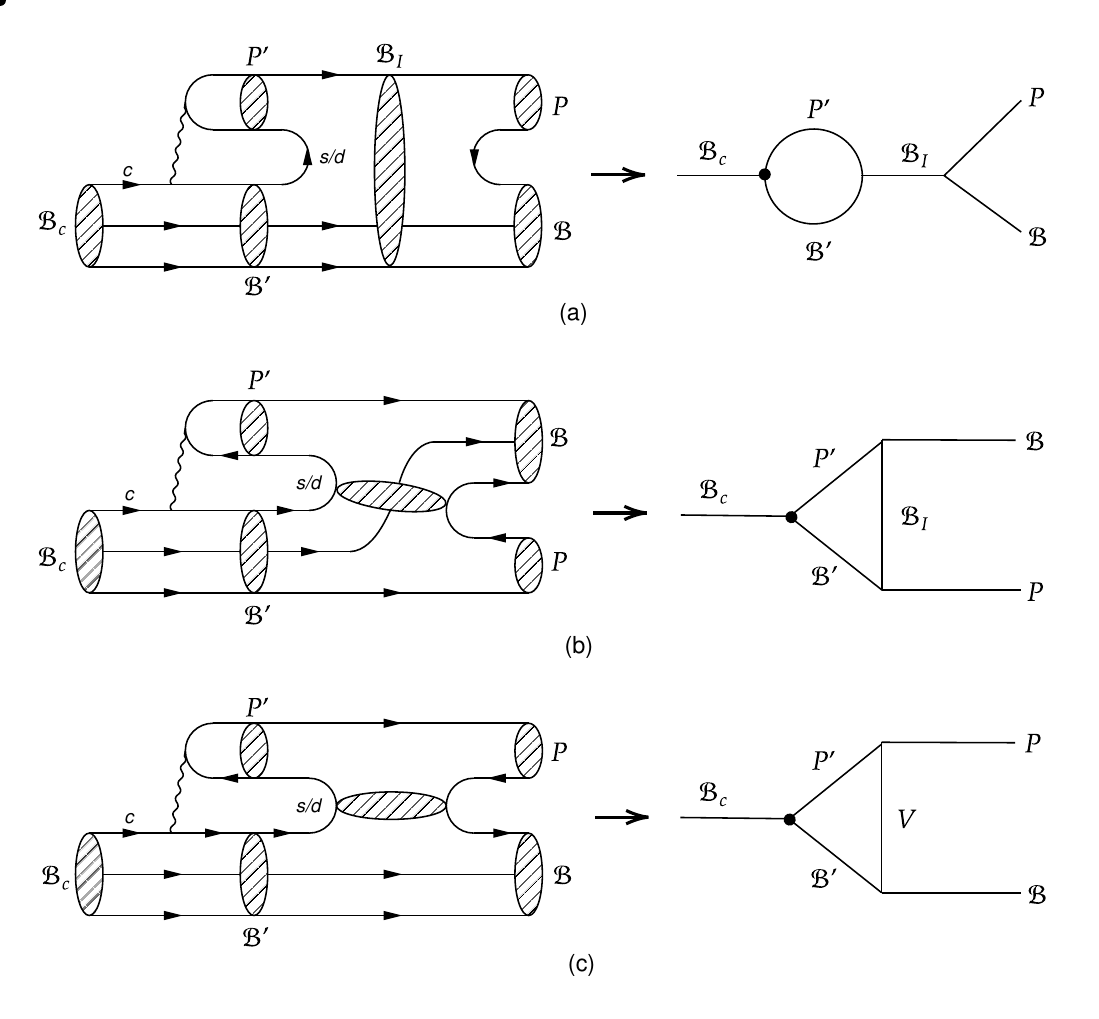}
\caption{\small Final-state $s$-, $t$- and $u$-channel rescattering contributions to $\B_c\to P\B$ from the external $W$-emission, depicted in (a), (b) and (c), respectively.  They have the same topology as the penguin one.
Pole and triangle diagrams are the corresponding processes at the hadron level. The blob at the weak vertex of the pole or triangle diagram represents SD contributions from the tree topologies $T$ and $C$.  
}
\label{Fig:FSI_baryon}
\end{figure}

\subsubsection{Final-state rescattering}
Our next task is to determine the coefficients $\tilde f^i_3$ in the IRA (see Eq. (\ref{eq:IRAHe})) or $\tilde b^i$ in the TDA, Eq. (\ref{eq:TDAlambdab2}) in order to study \CP violation in the charmed baryon sector. In analogue to the charmed meson case, the penguin topology can be generated through final-state $s$-, $t$- and $u$-channel rescattering
contributions to $\B_c\to P\B$ from the external $W$-emission as depicted in Fig. \ref{Fig:FSI_baryon}. Similar diagrams can be drawn for the internal $W$-emission $C$. Therefore, the blob at the weak vertex of the pole or triangle diagram represents SD contributions from the tree topologies $T$ and $C$.  
Although we do not know how to evaluate these diagrams at the quark level, their contributions at the hadron level manifested as the pole and triangle diagrams are amenable. For our purpose we will follow the work of He and Liu \cite{He:2024pxh} closely. 

For the SD weak vertex of $\B_c\to \B P$, we notice that
\begin{equation}
\begin{split}
\la \B P|{\cal L}_{\B_c \B P}|\B_c\ra & ={G_F\over\sqrt{2}}\left[ c_+(H_+)^{ij}_k +c_-(H_-)^{ij}_k\right]\la P|(\bar q_j^\alpha q^k_\alpha)|0\ra\la\B|(\bar q_i^\beta c_\beta)|\B_c\ra \\
& +{G_F\over\sqrt{2}}\left[ c_+(H_+)^{ij}_k -c_-(H_-)^{ij}_k\right]\la P |(\bar q_j^\beta q^k_\alpha)|0\ra\la\B|(\bar q_i^\alpha c_\beta)|\B_c\ra \\
\end{split}
\end{equation} 
and hence
\begin{equation}
\label{eq:BcBP}
{\cal L}_{\B_c \B P}=\sum_{P,\B, \B_c}F_{\B_c\B P}P^\dagger \ov \B\B_c= (P^\dagger)^k_j (\ov\B)^l_i\left( \tilde F_V^+(H_+)^{ij}_k+\tilde F_V^-(H_-)^{ij}_k \right) (\B_c)_l, 
\end{equation} 
where $H_+$ and $H_-$ are given in Eq. (\ref{eq:Hp,Hm}). In the above equation, the parameters $\tilde F^+_V$ and 
$\tilde F^-_V$ represent the overall unknowns introduced in Ref.~\cite{He:2024pxh} . We shall see later that $\tilde F^\pm_V$  is identical to $\tilde T\pm \tilde C$ in the TDA. The strong vertex of $\B\B P$ needed in $s$- and $t$-channel diagrams is given by
\be
{\cal L}_{\B \B' P}&=&\sum_{\B',P}\left[\sum_{\B_-} g_{\B_-\B' P}P\ov\B'\B_- + \sum_{\B_+} g_{\B_+\B' P}P\ov \B' i\gamma_5\B_+\right]  \non \\ 
&=& g_-\left[ (P)^i_j(\ov \B')^j_k(B_-)^k_i+ r_-(P)^j_k(\ov \B')^i_j(B_-)^k_i\right] \\
&+&  g_+\left[ (P)^i_j(\ov\B')^j_ki\gamma_5(B_+)^k_i+ r_+(P)^j_k(\ov \B')^i_j i\gamma_5(B_+)^k_i\right],  \non 
\en
where the subscript $\pm$ denotes the parity of $\B_\pm$.  The strong couplings $g_{\B_\pm\B' P}$ are related to 
$g_\pm$ and $r_\pm$ (see also Appendix A).

We next turn to the interaction of the vector meson with octet baryons. We shall assume ideal mixing of             $\omega$ and $\phi$ so that $\phi=s\bar s$ and $\omega=(u\bar u+d\bar d)/\sqrt{2}$. The vector meson matrix
reads
\begin{equation}
\label{eq:Vmatrix}
 V^{i}_{j}=\left(\begin{array}{ccc}
\frac{\rho^0+ \omega}{\sqrt{2}} & \rho^+ & K^{*+} \\
\rho^- & \frac{-\rho^0+ \omega}{\sqrt{2}} & K^{*0} \\
K^{*-} & \overline{K}^{*0} & \phi
\end{array}\right).
\end{equation}
The relevant $V\!PP$ interaction is given by
\be
{\cal L}_{V\!PP} &=& g_{V\!PP}{\rm Tr}V_\mu [P,\partial^\mu P] \non \\
  &=&
 -ig_{V\!PP}\Big(\rho^+_\mu\pi^0\!\lrpartial{}^{\!\mu}\pi^-
 +\rho^-_\mu\pi^+\!\lrpartial{}^{\!\mu}\pi^0+\rho^0_\mu\pi^-\!\lrpartial{}^{\!\mu}\pi^+ \non \\
 &&+ K^{*-}_\mu K^+\!\lrpartial{}^{\!\mu}\pi^0 +\sqrt{3}K^{*-}_\mu K^+\!\lrpartial{}^{\!\mu}\eta +\sqrt{3}(\phi_8)_\mu K^+\!\lrpartial{}^{\!\mu}K^- +\cdots\Big),
 \en
 where $\phi_8$ is referred to the octet component of the $\phi$ meson, which is given by $\phi\cos\theta$ with $\theta=35.3^\circ$  being the ideal $\omega-\phi$ mixing angle \cite{Cheng:2011fk}.
Applying  ${\cal L}_{V\!PP}$ to $V\to P_1P_2$ decays leads to
\be
g_{\rho^+\to \pi^+\pi^0} &=&  2g_V, \qquad
g_{K^{*+}\to K^+ \pi^0} =  g_V, \qquad
g_{\phi\to K^+  K^-} =  g_V\sqrt{3}\cos\phi=1.42 g_V,
\en
where the abbreviation of $g_{V\!PP}=g_V$ has been made.
Experimentally, $g_{\rho^+\to \pi^+\pi^0} =5.94$, $g_{K^{*+}\to K^+ \pi^0}=3.14$ and $g_{\phi\to K^+ K^-}=4.47$. The small discrepancy between theory and experiment is presumably ascribed to SU(3) breaking. 
The interaction of the vector meson with octet baryons is given by
\be
\label{eq:f1,f2}
{\cal L}_{V\!\B'\B} &=& f_{V\B'\B}{\rm Tr}(\ov\B' \gamma_\mu  V^\mu\B)+{ \tilde f_{V\!\B'\B}\over m_{\B'}+m_{\B} }
{\rm Tr}(\ov\B' \sigma_{\mu\nu}\B \,\partial^\mu V^\nu).
\en
The vector and tensor couplings $f_{V\B\B}$ and $\tilde f_{V\B\B}$, respectively,  have been evaluated using QCD sum rules \cite{Aliev:2009}.  

Just like the $P\B \B$ couplings, we can write
\be
\label{eq:f1}
\begin{split}
&
 f_{n \Sigma^- K^{*+}} = f_+, \quad f_{\Lambda p K^{*+}} = \frac{f_+}{\sqrt{6}}(1-2\rho_+), \\
&
 f_{\Lambda \Sigma^- \rho^+} = \frac{f_+}{\sqrt{6}}(1+\rho_+),\quad
 f_{\Sigma^0 \Sigma^+ \rho^-}=\frac{f_{+}}{\sqrt{2}}(1-\rho_+),
\end{split}
\en
and
\be
\label{eq:f2}
\begin{split}
&
 \tilde f_{n \Sigma^- K^{*+}} =\tilde f_+, \quad \tilde f_{\Lambda p K^{*+}} = \frac{\tilde f_+}{\sqrt{6}}(1-2\rho_+), \\
&
 \tilde f_{\Lambda \Sigma^- \tilde \rho^+} = \frac{\tilde f_+}{\sqrt{6}}(1+\tilde \rho_+),\quad
 \tilde f_{\Sigma^0 \Sigma^+ \tilde \rho^-}=\frac{\tilde f_{+}}{\sqrt{2}}(1-\tilde \rho_+)
\end{split}
\en
for baryons with positive parity. 


Since the $s$- and $t$-channel contributions have been evaluated by He and Liu Ref.~\cite{He:2024pxh} , we will focus on the $u$-channel contribution in Fig. \ref{Fig:FSI_baryon}(c) which was not considered by them. The relevant amplitude reads
\begin{equation}
\begin{split}
\label{eq:u}
A^u& =\sum_{V,\B',P'}\bar u_\B\left( \int {d^4 q\over (2\pi)^4}\,\left[f_{V\!\B \B'}\gamma_\mu+{\tilde f_{V\!\B\B'}\over m_{\B'}+m_{\B} }i\sigma_{\mu\nu}q^\nu\right] g_{V\! P P'}
(p_1+p_3)_\alpha(p\!\!\!/_2+m_{\B'}) \right. \\
& \qquad\qquad \times \left.    \Big(g^{\mu\alpha}-{q^\mu q^\alpha\over m_V^2}\Big) {F_{\B_c\B' P'} \over (q^2-m_V^2)(p_1^2-m_{P'}^2)(p_2^2-m_{\B'}^2) } \right) u_{\B_c}, \\
& =\sum_{V,\B',P'}\bar u_\B\left( \int {d^4 q\over (2\pi)^4}\,\left[f_{V\!\B \B'}\left( 2p\!\!\!/_3+q\!\!\!/
{m_V^2-q^2-2q\cdot p_3\over m_V^2}\right) +{\tilde f_{V\!\B\B'}\over m_{\B'}+m_{\B} }(q\!\!\!/p\!\!\!/_3-p\!\!\!/_3 q\!\!\!/)   \right] \right. \\
& \qquad\qquad \times \left.     {  g_{V\! P P'}  F_{\B_c\B' P'} (p\!\!\!/_4-q\!\!\!/ +m_{\B'}) \over (q^2-m_V^2)(q^2+2p_3\cdot q+m_P^2-m_{P'}^2)(q^2-2p_4\cdot q+m_{\B}^2-m_{\B'}^2) } \right) u_{\B_c}, \\
\end{split}
\end{equation} 
where  the momenta of $P', \B', P, \B$ and $V$ are denoted by $p_1, p_2, p_3, p_4$ and $q$, respectively. 
The $u$-channel amplitude 
\be
A^u=\sum_{\B',P',V}\left\la {\cal L}_{V\! PP'} {\cal L}_{V\!\B\B'} {\cal L}_{\B_c \B' P'}\right\ra
\en
will have the same expression as Eq. (\ref{eq:u}) after the contractions of $\B'\bar \B'$, $P' P'^\dagger$ and $VV^\dagger$ 
into the hadron propagators.  However, unlike the $s$- or $t$-channel amplitude which has the expression
\be
A^t=\bar u_\B\left[ \int {d^4 q\over (2\pi)^4}\Bigg( \sum_{\B_I,\B',P'} g_{\B_I^-\B P}g_{\B_I \B' P'}F_{\B_c\B' P'}\Bigg) I(q^2)\right]u_{\B_c},
\en
where $I(q^2)$ is flavor-independent, 
the $u$-channel amplitude cannot be expressed in the same simple form owing to the existence of both vector and tensor couplings, $f_{V\!\B \B'}$ and $\tilde f_{V\!\B \B'}$. In order to avoid the divergence occurred in the loop integral in Eq. (\ref{eq:u}), it is necessary to introduce some cutoffs on the couplings such as $f_{V\!\B \B'}$, $\tilde f_{V\!\B \B'}$, $g_{V\! P P'}$ and the form factor  $F_{\B_c\B' P'}$. The results depend on the detail of the cutoff mechanism.

The vector and tensor couplings have been evaluated using light cone QCD sum rules. From Tables I and II of Ref. \cite{Aliev:2009} for the numerical values of 
$f_{V\B'\B}$ and $\tilde f_{V\B'\B}$ (denoted by $f_1$ and $f_2$ in \cite{Aliev:2009}, respectively), we
obtain $f_+=6.1, \rho_+=-0.38, \tilde{f}_+=-1.6$ and $\tilde \rho_+ =-21$, where $f_+, \rho_+, \tilde{f}_+$ and $\tilde \rho_+ $ are defined in Eqs. (\ref{eq:f1}) and (\ref{eq:f2}).  In general, the tensor coupling is larger than the vector one by one order of magnitude, $|\tilde f_{V\B'\B}/ f_{V\B'\B}|\approx 11$. 
If we recast the $u$-channel contribution to the form
\be
A^u = \bar u_\B \left[ a f_{V\B'\B} + b \tilde f_{V\B'\B}\right] u_{\B_c},
\en
the ratio of $b/a$ will depend on the cutoff mechanism and the decay mode. Take the decay $\Lambda_c^+ \to p \phi$ as an example. 
An explicit triangle-diagram calculation of this mode in the framework of Ref.~\cite{Jia:2024pyb} with $\Lambda, K^+, K^{*+}$ involved in the loop yields 
$|b/a|\approx 1.06$ \cite{Jia}. The cutoff mechanism is given in Eq. (\ref{eq:cutoff}) below and will be discussed over there.
This indicates that the tensor  coupling gives a dominant contribution to the decay $\Lambda_c^+ \to p \phi$. To avoid the complication caused by the vector meson interaction with  baryons, we shall assume hereafter that the $u$-channel decay amplitude of $\B_c\to P\B$ mediated by the vector-meson exchange is dominated by the tensor coupling $\tilde f_{V\B'\B}$.

We now focus on the flavor structure of the $u$-channel amplitude 
\begin {equation}
\begin{split}
\label{eq:Au}
A^u 
 \propto \sum_{\B',P',V} (P')^{i_1}_{j_1}(P^\dagger)^{j_1}_{k_1} (V)^{k_1}_{i_1}(V^\dagger)^{j_2}_{k_2} 
 \left( (\ov\B)^{i_2}_{j_2}(\B')^{k_2}_{i_2}+\rho_+ (\ov\B)^{k_2}_{i_2}(\B')^{i_2}_{j_2}\right)  \left( (P'^\dagger)^k_i(\ov\B')^l_j(H_-)^{ij}_k(\B_c)_l\right),
\end{split}
\end{equation} 
where we have switched the notation from $\tilde \rho_+$ to $\rho_+$ for the reason of simplicity.
It was pointed out in Ref.~\cite{Liu:2023dvg} that the summation of the intermediate states in Eq. (\ref{eq:Au})  can be evaluated in the following manner. Writing 
\be
\label{eq:octet}
\B'={1\over\sqrt{2}}\sum_{a=1}^8\B'_a \lambda_a, \qquad P'={1\over\sqrt{2}}\sum_{a=1}^8 P'_a \lambda_a, \qquad
V={1\over\sqrt{2}}\sum_{a=0}^8 V_a \lambda_a, 
\en
under SU(3) flavor symmetry, where $\lambda_a$ are the Gell-Mann matrices, one can get ride of the Gell-Mann matrices by the completeness relation
\begin{equation}
\label{eq:complete}
\sum_{a=1}^8(\lambda_a)^i_j(\lambda_a)^k_l  =2\left(\delta^i_l\delta^k_j-\frac13\delta^i_j\delta^k_l\right).
\end{equation} 
However,  for the vector mesons
we shall use
\begin {equation}
\label{eq:completeV}
\sum_{a=0}^8(\lambda_a)^i_j(\lambda_a)^k_l =(\lambda_0)^i_j(\lambda_0)^k_l+\sum_{a=1}^8(\lambda_a)^i_j(\lambda_a)^k_l
=2\delta^i_l\delta^k_j.  \\
\end{equation} 
because of the existence of a nonet for the vector mesons, see Eq. (\ref{eq:Vmatrix}). For the pseudoscalar mesons, it is known that the SU(3) singlet $\eta_1$ acquires its mass through the axial anomaly. Therefore, we should not consider the $\eta_1$ meson inside the loop. In other words, we use the matrix $P'$ in Eq. (\ref{eq:octet}) for the pseudoscalar 
meson inside the loop and the matrix $P$ in Eq. (\ref{eq:B&P}) for the emitted physical meson. The matrix $P'$ is the same as $P$ with the $\eta_1$ terms removed.

After applying Eqs. (\ref{eq:octet}), (\ref{eq:complete}) and (\ref{eq:completeV}) to Eq. (\ref{eq:Au}), we  obtain
\begin {equation}
\begin{split}
A^u 
  & \propto ~ a_4 - \frac12(P)^{k}_{k}(\B)^{l}_{i}(H_{3})^{i}(\B_c)_l
  +\frac12(P)^{i}_{j}(\B)^{l}_{i}(H_{3})^{j}(\B_c)_l \\
  & ~~ -\frac13(1+4\rho_+)\left[ -a_2 -\frac12(P)^{j}_{l}(\B)^{l}_{i}(H_{3})^{i}(\B_c)_j 
  +\frac12(P)^{i}_{l}(\B)^{l}_{i}(H_{3})^{j}(\B_c)_j   \right] \\
  & ~~ - \frac13 \rho_+ (P)^{l}_{k}(\B)^{k}_{i}(H_{3})^{i}(\B_c)_l  -\frac13 \left(1+\rho_+\right)(P)^{k}_{i}(\B)^{l}_{k}(H_{3})^{i}(\B_c)_l \\
  & ~~ + \frac19(1+\rho_+) (P)^{l}_{k}(\B)^{k}_{l}(H_{3})^{i}(\B_c)_i ,
\end{split}
\end{equation} 
where $a_i$ are those terms in Eq. (\ref{eq:IRAamp}). Using the relations
\begin{equation}
\tilde f^a=-a_1+a_5, \qquad \tilde f^b=-a_2+a_4, \qquad \tilde f^c=-a_3-a_5, \qquad \tilde f^d=a_4+a_5,
\end{equation}
inferred from Eq. (\ref{eq:tildeTDAIRA2}),
we find
\begin {equation}
\begin{split}
A^u 
     &=~ U^-\left[ \frac23(1-2\rho_+)\tilde f^b  +\tilde f^d  -\frac12\tilde f^a_3 
  +\frac16(1+2\rho_+ )\tilde f^b_3\right. \\
  & \left.-\frac1{18}(1+10\rho_+)\tilde f^c_3+\frac16(1-2\rho_+)\tilde f^d_3 \right],
\end{split}
\end{equation} 
where we have assigned an overall unknown $U^-$. 

The $s$- and $t$-channel rescattering amplitudes have been evaluated in Ref. \cite{He:2024pxh} which we are able to 
reproduce. The results read
\begin{equation}
\begin{split}
\label{}
A^s &=S^-\left[ -\frac13(r_-+4)\tilde f^b-\frac13( r_-^2+4r_-) \tilde f^c -\frac16 (7r_--2)\tilde f^b_3 \right.\\
& +\left. \frac1{18}(r_- +1)(7r_--2)\tilde f^c_3-\frac16 r_-(7r_--2)\tilde f^d_3 \right],
\end{split}
\end{equation}
and 
\begin {equation}
\begin{split}
\label{eq:At}
A^t              
& ~~= \sum_{\lambda=\pm}T_\lambda^-\left[-\frac19 (2r_\lambda^2+4r_\lambda+11)\tilde f^a +\frac13(2r_\lambda^2-r_\lambda)\tilde f^b +\frac13(r_\lambda^2-2r_\lambda+3)\tilde f^c \right.\\
& ~~~\left. +\frac13(2r_\lambda^2-2r_\lambda -4)\tilde f^d  -\frac1{18}(10r_\lambda^2+2r_\lambda+1) \tilde f^a_3 +\frac13(r_\lambda^2-\frac52 r_\lambda+1)\tilde f^b_3   \right. \\
& ~~~ \left. -\frac1{18}(r_\lambda^2+11r_\lambda+1)\tilde f^c_3
+\frac16(r_\lambda+1)^2\tilde f^d_3\right],
\end{split}
\end{equation} 
where we have taken the contributions from the intermediate state $\B_I$ with both positive and negative parities into account.
Note that the hairpin contributions $\tilde f^a$ and $\tilde f^a_3$ in $A^t$ were not considered in Ref. \cite{He:2024pxh} as $P^i_i=\sqrt{3}\eta_1$.  In our treatment, although the SU(3)-singlet $\eta_1$ is not considered
for the pseudoscalar meson $P'$ in the loop, its presence is allowed for the emitted one $P$.

Finally, we consider the factorizable contributions to $\B_c\to \B P$ decays. In the TDA, it arise from the topological diagrams $T$ and $C$. This has been done before in Eqs. (\ref{eq:tildeTDAIRA2}) and (\ref{eq:bprime}): 
\begin {equation}
\begin{split}
& \tilde f^b=T- C, \qquad \tilde f^d= T- C, \qquad \tilde f^e= T+ C,  \\
& \tilde f^a_3=\frac14(T-3C), \qquad \tilde f^d_3=-\frac14(3T- C).
\end{split}
\end{equation} 
where we have dropped nonfactorizable terms such as $E_{1S}$, $E_{1A}$, $E_3$ and $E_h$. If we employ Eq. (\ref{eq:BcBP}) to evaluate the factorizable contributions, we will obtain \cite{He:2024pxh} 
\begin {equation}
\begin{split}
& \tilde f^b=\tilde{F}_V^- \qquad \tilde f^d= \tilde{F}_V^-, \qquad \tilde f^e= \tilde{F}_V^+,  \\
& \tilde f^a_3=\frac14(-\tilde{F}_V^++2\tilde{F}_V^-), \qquad \tilde f^d_3=-\frac14(\tilde{F}_V^+ +2\tilde{F}_V^- ).
\end{split}
\end{equation} 
Therefore, $\tilde F_V^\pm$ in Eq. (\ref{eq:BcBP}) correspond to $\tilde T\pm\tilde C$ in the TDA.

Collecting all the contributions from $A^{\rm fact}$ and $A^{s,t,u}$, we have
\begin {equation}
\begin{split}
\label{}
\tilde f^a & = 
\frac13\sum_{\lambda=\pm}(2r_\lambda^2+4r_\lambda+11)T^-_\lambda,  \\
\tilde f^b & = \tilde F_V^- -(r_-+4)S^- - 2(2\rho_+ -1)U^-+\sum_{\lambda=\pm}(2r_\lambda^2-r_\lambda)T^-_\lambda , \\
\tilde f^c & = -r_-(r_-+4)S^- 
+\sum_{\lambda=\pm}(r_\lambda^2-2r_\lambda+3)T^-_\lambda, \\
\tilde f^d & = \tilde F_V^-  +3U^-
+ \sum_{\lambda=\pm}(2r_\lambda^2-2r_\lambda-4)T^-_\lambda , \\
\tilde f^e & = \tilde F_V^+, \\
\end{split}
\end{equation} 
and
\begin {equation}
\begin{split}
\label{}
\tilde f^a_3 & =  \frac14(-\tilde F^+_V+2\tilde F_V^-) -\frac32 U^- -\frac16\sum_{\lambda=\pm}(10r_\lambda^2+2r_\lambda+1)T^-_\lambda , \\
\tilde f^b_3 & = -\frac12 (7r_--2)S^- +\frac12 (2\rho_+ +1) U^- +\frac12\sum_{\lambda=\pm}(2r_\lambda^2-5 r_\lambda+2)T^-_\lambda, \\
\tilde f^c_3 & = \frac16(r_-+1)(7r_--2)S^- - \frac16(10\rho_+ +1)U^- - \frac16\sum_{\lambda=\pm}(r_\lambda^2+11r_\lambda+1)T^-_\lambda  , \\
\tilde f^d_3 & = -\frac14(\tilde F^+_V+2\tilde F_V^-) -\frac12 r_-(7r_--2)S^- +\frac12 U^- +\frac12\sum_{\lambda=\pm}(r_\lambda+1)^2 T^-_\lambda.
\end{split}
\end{equation} 

If we follow Ref.~\cite{He:2024pxh}  to use $r_-=r_+=2.5\pm0.8$ and define $T^-\equiv T^-_+ + T^-_-$, then we have
\begin {equation}
\begin{split}
\label{eq:fFSI2}
\tilde f^a & = -\frac13(2r_-^2+4r_-+11)T^- \\
\tilde f^b & = \tilde F_V^- -(r_-+4)S^- - 2(2\rho_+-1)U^-+(2r_-^2-r_-)T^- , \\
\tilde f^c & =  -r_-(r_-+4)S^- 
+ (r_-^2-2r_-+3)T^-, \\
\tilde f^d & =\tilde F_V^- + 3 U^- + (2r_-^2-2r_--4)T^-, \\
\tilde f^e & = \tilde F_V^+, \\
\tilde f^a_3 & =  \frac14(-\tilde F^+_V+2\tilde F_V^-) -\frac32 U^- -\frac16(10r_-^2+2r_-+1)T^-, \\
\tilde f^b_3 & = -\frac12 (7r_--2)S^- +\frac12 (1+2\rho_+) U^- +\frac12(2r_-^2-5 r_-+2)T^-, \\
\tilde f^c_3 & = \frac16(r_-+1)(7r_--2)S^- - \frac16(r_-^2+11r_-+1)T^- - \frac16(10\rho_+ +1) U^- , \\
\tilde f^d_3 & = -\frac14(\tilde F^+_V+2\tilde F_V^-) -\frac12 r_-(7r_--2)S^- +\frac12 U^- +\frac12(r_-+1)^2 T^-,
\end{split}
\end{equation} 
in the IRA.  
The naturally established relation between the IRA coefficients and the FSR parameters arises from the fact that the FSR calculations are performed within the IRA framework.
Using the relations between TDA and IRA given in Eq. (\ref{eq:tildeTDAIRA1}) we obtain
\begin {equation}
\begin{split}
\label{}
\tilde T & = \frac12 (\tilde F_V^+ +\tilde F_V^-) -\frac12(r_-+4)S^- +\frac12(2r_-^2-r_-)T^- - (2\rho_+ -1)U^-, \\
\tilde C & = \frac12 (\tilde F_V^+ -\tilde F_V^-) +\frac12(r_-+4)S^- -\frac12(2r_-^2-r_-)T^- + (2\rho_+ -1)U^-, \\
\tilde C' & = -(r_-+4)S^- + (r_-+4)T^- - ( 4\rho_+ +1) U^-, \\
\tilde E_1 & = r_-(r_-+4)S^--(r_-^2-2r_-+3)T^-, \\
\tilde E_h & = -\frac13(2r_-^2+4r_-+11)T^- , \\
\end{split}
\label{eq:TDAtree}
\end{equation} 
and
\begin {equation}
\begin{split}
\label{eq:biFSI}
\tilde b_1 & =  \frac14(-\tilde F^+_V+2\tilde F_V^-) -\frac16(10r_-^2+2r_-+1)T^- -\frac32 U^-, \\
\tilde b_2 & = -\frac12 (7r_--2)S^- +\frac12(2r_-^2-5 r_-+2)T^-+\frac12 (2\rho_+ +1) U^-, \\
\tilde b_3 & = \frac16(r_-+1)(7r_--2)S^- - \frac16(r_-^2+11r_-+1)T^-  
- \frac16(10\rho_+  +1)U^-, \\
\tilde b_4 & = -\frac14(\tilde F^+_V+2\tilde F_V^-) -\frac12 r_-(7r_--2)S^- +\frac12(r_-+1)^2 T^- +\frac12 U^-,
\end{split}
\end{equation} 
in the TDA.


\section{Numerical analysis: results and discussions}
\label{sec:Num}

\subsection{Input parameters}
 
 In our previous analysis \cite{Zhong:2024qqs}, the tree-level TDA coefficients were extracted through a global fit to the available 30 measurements (see Table~VII therein). The remaining $Z_2$ ambiguity in the relative phase shift, identified in Refs.~\cite{Zhong:2024qqs, Geng:2023pkr}, was subsequently resolved in the updated global fit presented in~\cite{Cheng:2024lsn}, which incorporated 38 measurements (see Table~XI in Ref.~\cite{Cheng:2024lsn}). This refinement was achieved by including the measured Lee--Yang parameter $\beta$ from the $\Lambda_c^+ \to \Lambda^0 \pi^+$ and 
 $\Lambda_c^+ \to \Lambda^0 K^+$ decay modes reported by LHCb~\cite{LHCb:2024tnq}. Additionally, the measured value of $\gamma$ in~\cite{LHCb:2024tnq} was also utilized in~\cite{Cheng:2024lsn} to clarify the relative magnitude of the $S$- and $P$-wave amplitudes, as indicated by the two sets of partial wave amplitudes observed in the $\Lambda_c^+ \to \Xi^0 K^+$ decay measured by BESIII~\cite{BESIII:2023wrw}.

In principle, the accumulation of more data allows for the extraction of increasingly precise information.  
By the end of 2024, three new decay modes of $\Xi_c^+$ have been measured by the Belle and Belle-II collaborations~\cite{Belle:2024xcs}, yielding:
\begin{equation}
\begin{split}
& \mathcal{B}(\Xi_c^+ \to p\, K_S^0) = (7.16 \pm 3.25) \times 10^{-4}, \\
& \mathcal{B}(\Xi_c^+ \to \Lambda\, \pi^+) = (4.52 \pm 2.09) \times 10^{-4}, \\
& \mathcal{B}(\Xi_c^+ \to \Sigma^0\, \pi^+) = (1.20 \pm 0.55) \times 10^{-4}.
\end{split}
\end{equation}
More recently, in 2025, additional efforts on $\Xi_c^+$ decays have led to the measurements~\cite{Belle-II:2025klu}:
\begin{equation}
\begin{split}
& \mathcal{B}(\Xi_c^+ \to \Sigma^+\, K_S^0) = (1.94 \pm 0.90) \times 10^{-3}, \\
& \mathcal{B}(\Xi_c^+ \to \Xi^0\, \pi^+) = (7.19 \pm 3.23) \times 10^{-3}, \\
& \mathcal{B}(\Xi_c^+ \to \Xi^0\, K^+) = (4.9 \pm 2.3) \times 10^{-4}.
\end{split}
\end{equation}
These six newly measured branching fractions, together with the data listed in Table~XI of~\cite{Cheng:2024lsn}, are incorporated as inputs in the present global fit. A complete summary of the updated experimental data is presented in Table~\ref{tab:expdata} in Appendix~\ref{app:exp}.

In this work, we focus particularly on singly Cabibbo-suppressed (SCS) processes, as \CP violation in the SM occurs only in those modes. In terms of the five independent TDA tree amplitudes $\tilde T$, $\tilde C$, $\tilde C'$,   $\tilde E_1$  and $\tilde E_h$ in $\A_{\rm{TDA}}^{\rm{tree}}$ and the four independent penguin-related amplitudes $\tilde b_{1,2,3,4,5}$
in  $\A_{\rm{TDA}}^{\lambda_b}$,  the explicit decay amplitudes for all SCS modes are presented in Table~\ref{tab:TDAamp}.

\begin{table}[tp!]\footnotesize
\centering
\caption{The decay amplitudes of SCS processes in the TDA.}
\label{tab:TDAamp}
\begin{tabular}{lll}
\hline
Channel & ~~~~~~$\widetilde{\mathit{\rm TDA}}$ & \\
\hline
$\Lambda_c^+ \to \Lambda K^+$ &  
$\frac{1}{\sqrt{6}}[\frac{1}{2}(\lambda_d-\lambda_s)(4\Tilde{T}-\Tilde{C'}+2\Tilde{E_{1}})
+\lambda_b(\Tilde{b}_2-2\Tilde{b}_4+\frac{1}{2}\Tilde{b}_5)]$
\\

$\Lambda_c^+ \to \Sigma^0 K^+$ & 
$\frac{1}{\sqrt{2}}[\frac{1}{2}(\lambda_d-\lambda_s)\Tilde{C'}
+\lambda_b \Tilde{b}_2]$
\\

$\Lambda_c^+ \to \Sigma^+ K^0$ &
$\frac{1}{2}(\lambda_d-\lambda_s)\Tilde{C'}
+\lambda_b \Tilde{b}_2$
\\

$\Lambda_c^+ \to p \pi^0$ &
$\frac{1}{\sqrt{2}}[\frac{1}{2}(\lambda_d-\lambda_s)(-2\Tilde{C}-\Tilde{C'}-\Tilde{E_{1}})
+\lambda_b(\Tilde{b}_4+\frac{3}{4}\Tilde{b}_5)]$
\\

$\Lambda_c^+ \to p \eta_8$ &  
$\frac{1}{\sqrt{6}}[\frac{1}{2}(\lambda_d-\lambda_s)(6\Tilde{C}+\Tilde{C'}-\Tilde{E_{1}})
+\lambda_b(-2\Tilde{b}_2+\Tilde{b}_4+\frac{3}{4}\Tilde{b}_5)]$
\\

$\Lambda_c^+ \to p \eta_1$&
$\frac{1}{\sqrt{3}}[\frac{1}{2}(\lambda_d-\lambda_s)(\Tilde{C'}-\Tilde{E_{1}}+3\Tilde{E_{h}})
+\lambda_b(3\Tilde{b}_1+\Tilde{b}_2+\Tilde{b}_4)]$
\\

$\Lambda_c^+ \to n \pi^+$&
$\frac{1}{2}(\lambda_d-\lambda_s)(2\Tilde{T}-\Tilde{C'}-\Tilde{E_{1}})
+\lambda_b(\Tilde{b}_4-\frac{1}{4}\Tilde{b}_5)$
\\

$\Xi_c^0 \to \Lambda \pi^0$&
$\frac{1}{2\sqrt{3}}[\frac{1}{2}(\lambda_d-\lambda_s)(-2\Tilde{C}-2\Tilde{C'}+\Tilde{E_{1}})
+\lambda_b(\Tilde{b}_2+\Tilde{b}_4+\frac{3}{4}\Tilde{b}_5)]$
\\

$\Xi_c^0 \to \Lambda \eta_8$&
$\frac{1}{6}[\frac{1}{2}(\lambda_d-\lambda_s)(
6\Tilde{C}+3\Tilde{E_{1}}
)
+\lambda_b(\Tilde{b}_2+6\Tilde{b}_3+\Tilde{b}_4+\frac{3}{4}\Tilde{b}_5)]$
\\

$\Xi_c^0 \to \Lambda \eta_1$&
$\frac{1}{3\sqrt{2}}[\frac{1}{2}(\lambda_d-\lambda_s)(3\Tilde{C'}-3\Tilde{E_{1}}+9\Tilde{E_{h}})
+\lambda_b(3\Tilde{b}_1+\Tilde{b}_2+\Tilde{b}_4)]$
\\

$\Xi_c^0 \to \Sigma^0 \pi^0$&
$\frac{1}{2}[\frac{1}{2}(\lambda_d-\lambda_s)(-2\Tilde{C}-\Tilde{E_{1}})
+\lambda_b(\Tilde{b}_2+2\Tilde{b}_3+\Tilde{b}_4+\frac{3}{4}\Tilde{b}_5)]$
\\

$\Xi_c^0 \to \Sigma^0 \eta_8$&
$\frac{1}{2\sqrt{3}}[\frac{1}{2}(\lambda_d-\lambda_s)(6\Tilde{C}+2\Tilde{C'}+\Tilde{E_{1}})
+\lambda_b(\Tilde{b}_2+\Tilde{b}_4+\frac{3}{4}\Tilde{b}_5)]$
\\

$\Xi_c^0 \to \Sigma^0 \eta_1$&
$\frac{1}{\sqrt{6}}[\frac{1}{2}(\lambda_d-\lambda_s)(-\Tilde{C'}+\Tilde{E_{1}}-3\Tilde{E_{h}})
+\lambda_b(3\Tilde{b}_1+\Tilde{b}_2+\Tilde{b}_4)]$
\\

$\Xi_c^0 \to \Sigma^+ \pi^-$&
$\frac{1}{2}(\lambda_d-\lambda_s)(-\Tilde{E_{1}})
+\lambda_b(\Tilde{b}_2+\Tilde{b}_3)$
\\

$\Xi_c^0 \to \Sigma^- \pi^+$&
$\frac{1}{2}(\lambda_d-\lambda_s)(2\Tilde{T})
+\lambda_b(\Tilde{b}_3+\Tilde{b}_4-\frac{1}{4}\Tilde{b}_5)$
\\

$\Xi_c^0 \to \Xi^0 K^0$&
$\frac{1}{2}(\lambda_d-\lambda_s)(-\Tilde{C'}-\Tilde{E_{1}})
+\lambda_b \Tilde{b}_3$
\\

$\Xi_c^0 \to \Xi^- K^+$&
$\frac{1}{2}(\lambda_d-\lambda_s)(-2\Tilde{T})
+\lambda_b(\Tilde{b}_3+\Tilde{b}_4-\frac{1}{4}\Tilde{b}_5)$
\\

$\Xi_c^0 \to p K^-$&
$\frac{1}{2}(\lambda_d-\lambda_s) \Tilde{E_{1}}
+\lambda_b(\Tilde{b}_2+\Tilde{b}_3)$
\\

$\Xi_c^0 \to n \bar{K}^0$&
$\frac{1}{2}(\lambda_d-\lambda_s)(\Tilde{C'}+\Tilde{E_{1}})
+ \lambda_b \Tilde{b}_3$
\\

$\Xi_c^+ \to \Lambda \pi^+$&
$\frac{1}{\sqrt{6}}[\frac{1}{2}(\lambda_d-\lambda_s)(-2\Tilde{T}+2\Tilde{C'}-\Tilde{E_{1}})
+ \lambda_b(-\Tilde{b}_2-\Tilde{b}_4+\frac{1}{4}\Tilde{b}_5)]$
\\

$\Xi_c^+ \to \Sigma^0 \pi^+$&
$\frac{1}{\sqrt{2}}[\frac{1}{2}(\lambda_d-\lambda_s)(2\Tilde{T}+\Tilde{E_{1}})
+ \lambda_b(-\Tilde{b}_2+\Tilde{b}_4-\frac{1}{4}\Tilde{b}_5)]$
\\

$\Xi_c^+ \to \Sigma^+ \pi^0$&
$\frac{1}{\sqrt{2}}[\frac{1}{2}(\lambda_d-\lambda_s)(2\Tilde{C}-\Tilde{E_{1}})
+ \lambda_b(\Tilde{b}_2-\Tilde{b}_4-\frac{3}{4}\Tilde{b}_5)]$
\\

$\Xi_c^+ \to \Sigma^+ \eta_8$&
$\frac{1}{\sqrt{6}}[\frac{1}{2}(\lambda_d-\lambda_s)(-6\Tilde{C}-2\Tilde{C'}-\Tilde{E_{1}})
+\lambda_b(-\Tilde{b}_2-\Tilde{b}_4-\frac{3}{4}\Tilde{b}_5)]$
\\

$\Xi_c^+ \to \Sigma^+ \eta_1$&
$\frac{1}{\sqrt{3}}[\frac{1}{2}(\lambda_d-\lambda_s)(\Tilde{C'}-\Tilde{E_{1}}+3\Tilde{E_{h}})
+ \lambda_b(-3\Tilde{b}_1-\Tilde{b}_2-\Tilde{b}_4)]$
\\

$\Xi_c^+ \to \Xi^0 K^+$&
$\frac{1}{2}(\lambda_d-\lambda_s)(2\Tilde{T}-\Tilde{C'}-\Tilde{E_{1}})
+\lambda_b(-\Tilde{b}_4+\frac{1}{4}\Tilde{b}_5)$
\\

$\Xi_c^+ \to p \bar{K}^0$&
$\frac{1}{2}(\lambda_d-\lambda_s)\Tilde{C'}
+\lambda_b(-\Tilde{b}_2)$
\\
\hline
\end{tabular}
\end{table}

The five unknown parameters \((\tilde F_V^+, \tilde F_V^-)\) and \((S^-, T^-, U^-)\) represent final-state rescattering contributions from short-distance (SD) and long-distance (LD) effects, respectively. The LD contributions are modeled via final-state rescattering through the $s$-, $t$-, and $u$-channel processes.
The experimental results summarized in Table~\ref{tab:expdata}, along with the theoretical decay amplitudes listed in Table~\ref{tab:TDAamp}, serve as key inputs to the construction of the $\chi^2$ function.

Since there exist 5 independent tilde TDA amplitudes given in Eq. (\ref{eq:tildeTDA}), we have totally 19 unknown parameters to describe the magnitudes and the phases of the respective $S$- and $P$-wave amplitudes, namely, 
\begin{eqnarray}
\label{eq:19parameters}
&& |\tilde T|_Se^{i\delta_S^{\tilde T}}, \quad |\tilde C|_Se^{i\delta_S^{\tilde C}}, \quad |\tilde C'|_Se^{i\delta_S^{\tilde C'}},  \quad |\tilde E_{1}|_Se^{i\delta_S^{\tilde E_{1}}}, \quad |\tilde E_h|_Se^{i\delta_S^{\tilde E_h}},\nonumber \\
&& |\tilde T|_Pe^{i\delta_P^{\tilde T}}, \quad |\tilde C|_Pe^{i\delta_P^{\tilde C}}, \quad |\tilde C'|_Pe^{i\delta_P^{\tilde C'}}, \quad |\tilde E_{1}|_Pe^{i\delta_P^{\tilde E_{1}}}, \quad |\tilde E_h|_Pe^{i\delta_P^{\tilde E_h}},
\end{eqnarray}
collectively denoted by $|X_i|_Se^{i\delta^{X_i}_S}$ and $|X_i|_Pe^{i\delta^{X_i}_P}$,
where the subscripts $S$ and $P$ denote the $S$- and $P$-wave components of each TDA amplitude. 
Since there is an overall phase which can be omitted, we shall set $\delta_S^{\tilde T}=0$. 
Likewise, for the TDA coefficients $\tilde b_{1,2,3,4}$
in  $\A_{\rm{TDA}}^{\lambda_b}$, we have
\begin{eqnarray}
\label{eq:b1...5}
&& |\tilde b_1|_Se^{i\delta_S^{\tilde b_1}}, \quad  |\tilde b_2|_Se^{i\delta_S^{\tilde b_2}}, \quad |\tilde b_3 |_Se^{i\delta_S^{\tilde b_3}}, \quad~  |\tilde b_4|_Se^{i\delta_S^{\tilde b_4}},  \nonumber \\
&& |\tilde b_1|_Pe^{i\delta_P^{\tilde b_1}}, \quad |\tilde b_2|_Pe^{i\delta_P^{\tilde b_2}}, \quad |\tilde b_3|_Pe^{i\delta_P^{\tilde b_3}},  \quad |\tilde b_4|_Pe^{i\delta_P^{\tilde b_4}}.
\end{eqnarray}

\subsection{Numerical results for \CP violation}

\begin{table}[t!]
\caption{The TDA coefficients and FSR parameters. 
The upper entries are tree-level coefficients obtained from the global fit in Case I.
The middle ones are FSR parameters extracted from Eq.~\eqref{eq:TDAtree} 
 while the 
lower entries associated  with penguin diagrams are solved from Eq. ~\eqref{eq:biFSI}.  
In the calculation, we have adopted
$
\rho_+=-21$ and $r_-=2.5$.
} 
\label{tab:TDAampCoeff}
\vspace{-0.1cm}
\begin{center}
\renewcommand\arraystretch{1}
\begin{tabular}
{c| c r r r }
\hline \hline
&$|X_i|_S$&$|X_i|_P$~~~ &$\delta^{X_i}_{S}$~~~~~ &$\delta^{X_i}_{P}$~~~~ \\
&\multicolumn{2}{c}{$(10^{-2}G_{F}~{\rm GeV}^2)$}&\multicolumn{2}{c}{$(\text{in radian})$} \\
\hline 
$\Tilde{T}$&
~$4.31\pm0.11$&$12.11\pm0.31$&
-- ~~~~~~~&$2.39\pm0.04$\\
$\Tilde{C}$&
~$3.23\pm0.48$&$11.35\pm0.93$&
$3.10\pm0.11$&$-0.72\pm0.16$\\
$\Tilde{C'}$&
~$5.84\pm0.35$&$17.74\pm0.92$&
$0.02\pm0.04$&$2.27\pm0.11$\\
$\Tilde{E_{1}}$&
~$2.79\pm0.19$&$10.41\pm0.47$&
$-2.81\pm0.06$&$1.83\pm0.09$\\
$\Tilde{E_{h}}$&
~$4.30\pm0.50$&$13.26\pm1.83$&
$2.70\pm0.11$&$-1.85\pm0.20$\\
\hline
{$\Tilde{F}_V^+$}&
~$1.10\pm0.43$&$0.83\pm0.46$&
$0.13\pm0.35$&$2.01\pm1.37$\\
{$\Tilde{F}_V^-$}&
~$0.48\pm0.43$&$7.01\pm1.79$&
$0.69\pm1.02$&$-2.81\pm0.15$\\
{$S^-$}&
~$0.12\pm0.01$&$0.92\pm0.05$&
$-2.20\pm0.11$&$1.66\pm0.08$\\
{$T^-$}&
~$0.39\pm0.04$&$1.19\pm0.16$&
$-0.45\pm0.11$&$1.29\pm0.20$\\
{$U^-$}&
~$0.04\pm0.00$&$0.22\pm0.01$&
$0.17\pm0.08$&$2.43\pm0.08$\\
\hline 
$\Tilde{b}_1$&
~$4.57\pm0.78$&$16.04\pm2.43$&
$2.89\pm0.13$&$-2.00\pm0.18$\\
$\Tilde{b}_2$&
~$0.49\pm0.11$&$10.00\pm0.22$&
$1.30\pm0.25$&$-1.11\pm0.07$\\
$\Tilde{b}_3$&
~$1.38\pm0.30$&$10.89\pm1.13$&
$2.93\pm0.19$&$2.47\pm0.11$\\
$\Tilde{b}_4$&
~$3.16\pm0.37$&$11.94\pm0.93$&
$0.23\pm0.10$&$-0.96\pm0.15$\\
$\Tilde{b}_5$&
~$1.10\pm0.43$&$0.83\pm0.46$&
$0.13\pm0.35$&$2.01\pm1.37$\\

\hline \hline
\end{tabular}
\end{center}
\end{table}

\begin{table}[b!]
\caption{The fitted IRA coefficients under the same condition  as that of Table \ref{tab:TDAampCoeff}.
} 
\label{tab:IRAampCoeff}
\vspace{-0.1cm}
\begin{center}
\renewcommand\arraystretch{1}
\begin{tabular}
{c| c r r r }
\hline \hline
&$|X_i|_S$&$|X_i|_P$~~~ &$\delta^{X_i}_{S}$~~~~~ &$\delta^{X_i}_{P}$~~~~ \\
&\multicolumn{2}{c}{$(10^{-2}G_{F}~{\rm GeV}^2)$}&\multicolumn{2}{c}{$(\text{in radian})$} \\
\hline 
$\Tilde{f}^a$&
~$4.30\pm0.50$&$13.26\pm1.82$&
-- ~~~~~~~&$1.74\pm0.15$\\
$\Tilde{f}^b$&
~$7.54\pm0.55$&$23.45\pm1.11$&
$-2.71\pm0.10$&$-0.29\pm0.11$\\
$\Tilde{f}^c$&
~$2.79\pm0.19$&$10.41\pm0.47$&
$-2.36\pm0.12$&$2.28\pm0.15$\\
$\Tilde{f}^d$&
~$1.71\pm0.51$&$6.36\pm1.26$&
$-2.84\pm0.19$&$0.10\pm0.20$\\
$\Tilde{f}^e$&
~$1.10\pm0.45$&$0.83\pm0.46$&
$-2.57\pm0.43$&$-0.69\pm2.40$\\
\hline 
$\Tilde{f}^a_3$&
~$4.57\pm0.81$&$16.04\pm2.43$&
$2.89\pm0.04$&$-1.80\pm0.14$\\
$\Tilde{f}^b_3$&
~$0.49\pm0.11$&$10.00\pm0.23$&
$1.50\pm0.29$&$-0.91\pm0.12$\\
$\Tilde{f}^c_3$&
~$1.38\pm0.31$&$10.89\pm1.19$&
$3.13\pm0.13$&$2.67\pm0.15$\\
$\Tilde{f}^d_3$&
~$3.16\pm0.38$&$11.94\pm0.94$&
$0.43\pm0.11$&$-0.75\pm0.21$\\
$\Tilde{f}^e_3$&
~$1.10\pm0.45$&$0.83\pm0.46$&
$0.33\pm0.43$&$2.21\pm2.40$\\
\hline \hline
\end{tabular}
\end{center}
\end{table}

The five sets of tree-level TDA parameters \((\tilde T, \tilde C, \tilde C', \tilde E_1, \tilde E_h)\) can be extracted through a global fit to the experimental data, including branching fractions and Lee–Yang parameters. For our purpose, we shall consider two different fits: Case I without the recent Belle data on $\Xi_c^0\to \Xi^0 \pi^0, \Xi^o\eta^{(')}$  \cite{Belle-II:2024jql} and Case II with all the data included. The fitted results of tree TDA parameters in Case I are shown in Table~\ref{tab:TDAampCoeff}. As for
the penguin-related TDA parameters $\tilde b_1, \ldots, \tilde b_4$, they cannot be determined owing to the lack of 
experimental data on \CP violation in the charmed baryon sector. 
The observation that sizable penguin topology can be induced 
from final-state rescattering (FSR) allows us to establish the relations
between the tree-level and penguin-level TDA coefficients after introducing some FSR parameters $\tilde F_V^+, \tilde F_V^-, S^-, T^-, U^-$.  They are ready extracted from Eq. (\ref{eq:TDAtree}).
In the middle of Table~\ref{tab:TDAampCoeff}, we present the numerical results for these FSR parameters, including their magnitudes and phases. It follows that $|T^-|> |S^-|> |U^-|$ for both $S$- and $P$-wave transitions from a global fit to the data. Notice that the smallness of $U^-$ is compensated by the largeness of $\rho_+$. Also we see that for the real part of $P$-wave, $|T^-|> |U^-|> |S^-|$.
Hence, although $|U^-|$ is smaller than $|T^-|$ and $ |S^-|$, the $u$-channel contribution is not negligible. Moreover, it is not justified to neglect the $s$-channel contribution as argued in Ref.~\cite{Jia:2024pyb}. 
 The fitted parameters in the IRA are summarized in Table~\ref{tab:IRAampCoeff}, 
which provides an independent cross-check of the TDA calculations, 
in spite of the existence of relations such as Eq.~\eqref{eq:tildeTDAIRA1}.

\begin{table}
\begin{center}
\caption{The predicted CP observables calculated with $\rho_+=-21$ and $r_-=2.5$ in both
TDA (upper) and IRA (lower).}
\vspace{0.3cm}
\label{tab:CPobsv}
\resizebox{0.9\textwidth}{!} 
{
\begin{tabular}{lrrrrrr}
\hline
Channel & 
$10^3 \mathcal{B}$ & 
$10^4 A_{CP}$ & 
$10^4 A_{CP}^S$ & 
$10^4 A_{CP}^P$ &
$10^4 A_{\alpha}$ & 
$10^4 R_{\beta}$ \\
\hline
\multirow{2}{*}{$\Lambda_c^+\to\Lambda^0 K^+$}&$0.64\pm0.03$&$0.49\pm0.87$&$5.07\pm2.27$&$-0.30\pm0.83$&$6.89\pm0.82$&$-7.66\pm1.67$\\&$0.64\pm0.03$&$0.49\pm0.91$&$5.07\pm2.30$&$-0.30\pm0.87$&$6.89\pm0.82$&$-7.66\pm1.68$\\
\multirow{2}{*}{$\Lambda_c^+\to\Sigma^0 K^+$}&$0.39\pm0.02$&$-1.33\pm0.27$&$-1.04\pm0.25$&$-1.73\pm0.64$&$-4.68\pm0.97$&$2.93\pm0.50$\\&$0.39\pm0.02$&$-1.33\pm0.28$&$-1.04\pm0.25$&$-1.73\pm0.67$&$-4.68\pm0.98$&$2.93\pm0.50$\\
\multirow{2}{*}{$\Lambda_c^+\to\Sigma^+K_S$}&$0.39\pm0.02$&$-1.33\pm0.27$&$-1.04\pm0.25$&$-1.73\pm0.64$&$-4.68\pm0.97$&$2.93\pm0.50$\\&$0.39\pm0.02$&$-1.33\pm0.28$&$-1.04\pm0.25$&$-1.73\pm0.67$&$-4.68\pm0.98$&$2.93\pm0.50$\\
\multirow{2}{*}{$\Lambda_c^+\to p\pi^0$}&$0.19\pm0.03$&$-7.88\pm2.92$&$-0.86\pm4.46$&$-13.13\pm2.76$&$6.97\pm9.17$&$-0.78\pm2.09$\\&$0.19\pm0.03$&$-7.88\pm2.95$&$-0.86\pm4.57$&$-13.13\pm2.75$&$6.97\pm9.78$&$-0.78\pm2.21$\\
\multirow{2}{*}{$\Lambda_c^+\to p\eta$}&$1.63\pm0.09$&$2.55\pm0.23$&$2.55\pm0.58$&$2.54\pm0.24$&$0.45\pm0.31$&$-0.77\pm0.44$\\&$1.63\pm0.09$&$2.55\pm0.22$&$2.55\pm0.59$&$2.54\pm0.24$&$0.45\pm0.31$&$-0.77\pm0.44$\\
\multirow{2}{*}{$\Lambda_c^+\to p\eta'$}&$0.52\pm0.08$&$-14.40\pm1.26$&$-17.31\pm4.08$&$-13.57\pm1.78$&$-1.81\pm2.62$&$-0.69\pm2.33$\\&$0.52\pm0.08$&$-14.40\pm1.26$&$-17.31\pm4.06$&$-13.57\pm1.78$&$-1.81\pm2.76$&$-0.69\pm2.35$\\
\multirow{2}{*}{$\Lambda_c^+\to n\pi^+$}&$0.61\pm0.06$&$-5.16\pm0.90$&$-0.65\pm0.84$&$-17.36\pm2.48$&$-5.55\pm3.86$&$-2.99\pm1.43$\\&$0.61\pm0.06$&$-5.16\pm0.90$&$-0.65\pm0.86$&$-17.36\pm2.54$&$-5.55\pm3.89$&$-2.99\pm1.47$\\
\multirow{2}{*}{$\Xi_c^0 \rightarrow \Lambda^0 \pi^0 $}&$0.07\pm0.01$&$1.76\pm2.31$&$0.92\pm0.41$&$29.36\pm57.57$&$51.88\pm308.2$&$-23.28\pm31.01$\\&$0.07\pm0.02$&$1.76\pm2.32$&$0.92\pm0.41$&$29.36\pm57.67$&$51.88\pm69.69$&$-23.28\pm16.56$\\
\multirow{2}{*}{$\Xi_c^0 \rightarrow \Lambda^0 \eta$}&$0.40\pm0.05$&$3.97\pm0.79$&$1.79\pm0.55$&$9.15\pm2.95$&$-4.07\pm0.87$&$3.52\pm1.69$\\&$0.40\pm0.05$&$3.97\pm0.82$&$1.79\pm0.54$&$9.15\pm3.17$&$-4.07\pm0.88$&$3.52\pm1.71$\\
\multirow{2}{*}{$\Xi_c^0 \rightarrow \Lambda^0 \eta'$}&$0.63\pm0.08$&$-3.96\pm0.38$&$-4.29\pm1.00$&$-3.84\pm0.61$&$0.32\pm0.52$&$-0.51\pm0.52$\\&$0.63\pm0.08$&$-3.96\pm0.38$&$-4.29\pm1.01$&$-3.84\pm0.62$&$0.32\pm0.53$&$-0.51\pm0.52$\\
\multirow{2}{*}{$\Xi_c^0 \rightarrow \Sigma^0 \pi^0 $}&$0.34\pm0.03$&$-3.26\pm0.72$&$-2.55\pm0.68$&$-5.40\pm2.71$&$-0.40\pm0.58$&$-2.45\pm2.90$\\&$0.34\pm0.03$&$-3.26\pm0.74$&$-2.55\pm0.69$&$-5.40\pm2.85$&$-0.40\pm0.59$&$-2.45\pm3.07$\\
\multirow{2}{*}{$\Xi_c^0 \rightarrow \Sigma^0 \eta$}&$0.17\pm0.03$&$11.58\pm1.55$&$5.20\pm1.37$&$16.52\pm4.41$&$-0.60\pm2.46$&$-14.72\pm37.07$\\&$0.17\pm0.03$&$11.58\pm1.62$&$5.20\pm1.41$&$16.52\pm4.58$&$-0.60\pm2.64$&$-14.72\pm25.20$\\
\multirow{2}{*}{$\Xi_c^0 \rightarrow \Sigma^0 \eta'$}&$0.18\pm0.03$&$4.73\pm0.96$&$5.56\pm2.11$&$4.07\pm1.86$&$-1.29\pm1.72$&$0.54\pm0.48$\\&$0.18\pm0.03$&$4.73\pm0.96$&$5.56\pm2.17$&$4.07\pm1.98$&$-1.29\pm1.79$&$0.54\pm0.48$\\
\multirow{2}{*}{$\Xi_c^0 \rightarrow \Sigma^{+} \pi^{-}$}&$0.26\pm0.02$&$1.61\pm1.04$&$-5.19\pm1.02$&$5.64\pm1.44$&$-41.25\pm68.74$&$-1.19\pm0.62$\\&$0.26\pm0.02$&$1.61\pm1.07$&$-5.19\pm1.02$&$5.64\pm1.52$&$-41.26\pm361.3$&$-1.19\pm0.64$\\
\multirow{2}{*}{$\Xi_c^0 \rightarrow \Sigma^- \pi^{+}$}&$1.80\pm0.05$&$-1.63\pm0.46$&$-1.47\pm0.39$&$-1.80\pm0.68$&$-1.28\pm0.34$&$1.46\pm0.30$\\&$1.80\pm0.05$&$-1.63\pm0.47$&$-1.47\pm0.40$&$-1.80\pm0.70$&$-1.28\pm0.34$&$1.46\pm0.30$\\
\multirow{2}{*}{$\Xi_c^0 \rightarrow \Xi^0 K_{S / L}$}&$0.38\pm0.01$&$1.50\pm0.50$&$-0.18\pm1.31$&$1.81\pm0.59$&$-5.86\pm0.48$&$4.38\pm1.48$\\&$0.38\pm0.01$&$1.50\pm0.53$&$-0.18\pm1.32$&$1.81\pm0.62$&$-5.86\pm0.48$&$4.38\pm1.48$\\
\multirow{2}{*}{$\Xi_c^0\to\Xi^-K^+$}&$1.31\pm0.04$&$1.59\pm0.43$&$1.47\pm0.39$&$1.80\pm0.68$&$1.32\pm0.34$&$-1.43\pm0.32$\\&$1.31\pm0.04$&$1.59\pm0.44$&$1.47\pm0.40$&$1.80\pm0.70$&$1.32\pm0.34$&$-1.43\pm0.32$\\
\multirow{2}{*}{$\Xi_c^0 \rightarrow p K^{-}$}&$0.31\pm0.02$&$-2.61\pm1.11$&$5.19\pm1.02$&$-5.64\pm1.44$&$41.98\pm67.85$&$2.19\pm0.69$\\&$0.31\pm0.02$&$-2.61\pm1.16$&$5.19\pm1.02$&$-5.64\pm1.52$&$41.98\pm307.3$&$2.19\pm0.71$\\
\multirow{2}{*}{$\Xi_c^0 \rightarrow n K_{S / L}$}&$0.83\pm0.04$&$-1.67\pm0.54$&$0.18\pm1.31$&$-1.82\pm0.59$&$6.03\pm0.51$&$-4.21\pm1.52$\\&$0.83\pm0.04$&$-1.67\pm0.57$&$0.18\pm1.32$&$-1.81\pm0.62$&$6.02\pm0.51$&$-4.21\pm1.52$\\
\multirow{2}{*}{$\Xi_c^{+} \rightarrow \Lambda^0 \pi^{+}$}&$0.21\pm0.04$&$3.15\pm3.95$&$1.26\pm0.60$&$81.20\pm171.2$&$62.14\pm548.6$&$-53.44\pm76.39$\\&$0.21\pm0.04$&$3.15\pm4.09$&$1.26\pm0.59$&$81.20\pm91.49$&$62.15\pm180.3$&$-53.44\pm188.1$\\
\multirow{2}{*}{$\Xi_c^{+} \rightarrow \Sigma^0 \pi^{+}$}&$3.16\pm0.09$&$0.16\pm0.65$&$-1.32\pm0.81$&$0.56\pm0.73$&$-3.66\pm0.46$&$2.85\pm0.47$\\&$3.16\pm0.09$&$0.16\pm0.68$&$-1.32\pm0.83$&$0.56\pm0.76$&$-3.66\pm0.46$&$2.85\pm0.47$\\
\multirow{2}{*}{$\Xi_c^{+} \rightarrow \Sigma^{+} \pi^0$}&$2.57\pm0.11$&$0.03\pm0.74$&$-4.80\pm2.96$&$0.65\pm0.68$&$-6.20\pm1.82$&$6.64\pm4.41$\\&$2.57\pm0.11$&$0.03\pm0.73$&$-4.80\pm2.96$&$0.65\pm0.68$&$-6.20\pm1.94$&$6.64\pm4.66$\\
\multirow{2}{*}{$\Xi_c^{+} \rightarrow \Sigma^{+} \eta$}&$1.02\pm0.17$&$11.59\pm1.55$&$5.20\pm1.37$&$16.52\pm4.41$&$-0.62\pm2.45$&$-14.73\pm37.07$\\&$1.02\pm0.16$&$11.59\pm1.62$&$5.20\pm1.41$&$16.52\pm4.58$&$-0.62\pm2.64$&$-14.73\pm25.21$\\
\multirow{2}{*}{$\Xi_c^{+} \rightarrow \Sigma^{+} \eta'$}&$1.09\pm0.17$&$4.73\pm0.96$&$5.56\pm2.11$&$4.07\pm1.86$&$-1.29\pm1.73$&$0.54\pm0.48$\\&$1.09\pm0.18$&$4.73\pm0.96$&$5.56\pm2.17$&$4.07\pm1.98$&$-1.29\pm1.79$&$0.54\pm0.48$\\
\multirow{2}{*}{$\Xi_c^{+} \rightarrow \Xi^0 K^{+}$}&$1.15\pm0.10$&$3.02\pm0.73$&$0.65\pm0.84$&$17.35\pm2.48$&$7.70\pm3.62$&$5.12\pm1.62$\\&$1.15\pm0.10$&$3.02\pm0.73$&$0.65\pm0.86$&$17.35\pm2.53$&$7.70\pm3.64$&$5.12\pm1.69$\\
\multirow{2}{*}{$\Xi_c^{+} \rightarrow p K_{S / L}$}&$1.52\pm0.08$&$-1.47\pm0.39$&$-1.04\pm0.25$&$-1.73\pm0.64$&$-4.54\pm1.07$&$3.07\pm0.59$\\&$1.52\pm0.08$&$-1.47\pm0.40$&$-1.04\pm0.25$&$-1.73\pm0.67$&$-4.54\pm1.09$&$3.07\pm0.59$\\
\hline
\end{tabular}
}
\end{center}
\end{table}

Next, the remaining parameters associated with penguin contributions, \((\tilde b_1, \ldots, \tilde b_5)\), can be obtained from Eq.~\eqref{eq:biFSI} with the help of the previously determined FSR parameters. Each of these penguin parameters contains four degrees of freedom, corresponding to the magnitudes and phases of the $S$- and $P$-wave amplitudes. 

With the calculated $\tilde b_i$ values and the tree-level coefficients at hand, it is ready to compute
the {\it CP}-violating observables as listed  in Table~\ref{tab:CPobsv}.  In general, \CP asymmetries are small of order $10^{-4}$ or even smaller. Nevertheless,  
\CP violation at the per mille level is found in the following decay modes:
\begin{equation*}
\Lambda_c \to p \pi^0,\qquad 
\Lambda_c \to p \eta',\qquad 
\Xi_c^0 \to \Sigma^0 \eta,\qquad 
\Xi_c^+ \to \Sigma^+ \eta,
\end{equation*}
with the predictions:
\begin{equation}
\begin{split}
\ACP(\Lambda_c \to p \pi^0) &=- (0.8 \pm 0.3) \times 10^{-3}, \quad 
\ACP(\Lambda_c \to p \eta') = (1.4 \pm 0.1) \times 10^{-3}, \\
\ACP(\Xi_c^0 \to \Sigma^0 \eta) &= (1.2 \pm 0.2) \times 10^{-3}, \quad 
\ACP(\Xi_c^+ \to \Sigma^+ \eta) = (1.2 \pm 0.2) \times 10^{-3}.
\end{split}
\label{eq:CPVnum}
\end{equation}
Likewise, the following modes have $P$-wave \CP asymmetries of order $10^{-3}$
\be
&& \Lambda_c^+ \to p \pi^0,\quad 
\Lambda_c^+ \to p \eta',\quad 
~\,\Lambda_c^+ \to n \pi^+,\quad 
\Xi_c^0 \to \Lambda \pi^0, \non \\
 && \Xi_c^0 \to \Sigma^0 \eta,\quad 
 \Xi_c^+ \to \Lambda \pi^+,\quad 
\Xi_c^+ \to \Sigma^+ \eta,\quad   \Xi_c^+ \to\Xi^0 K^+,
\en
while among them, only the decay $\Lambda_c^+ \to p \eta'$ has a large $S$-wave \CP asymmetry.
From Table~\ref{tab:CPobsv} we also see that
the $U$-spin symmetry relations \cite{He:2018joe,Wang:2019dls}
\begin{equation}
\begin{split}
\ACP(\Lambda_c^+ \to n \pi^+) & = -\ACP(\Xi_c^+ \to \Xi^0 K^+), \\
\ACP(\Lambda_c^+ \to \Sigma^+ K_{S,L}) & = -\ACP(\Xi_c^+ \to p K_{S.L}), \\
\ACP(\Xi_c^0 \to \Xi^0 K_{S,L}) &= -\ACP(\Xi_c^0 \to n K_{S,L}), \\
\ACP(\Xi_c^0 \to \Sigma^- \pi^+) & = -\ACP(\Xi_c^0 \to \Xi^- K^+), \\
\ACP(\Xi_c^0 \to \Sigma^+ \pi^-) & = -\ACP(\Xi_c^0 \to p K^-), \\
\end{split}
\label{eq:Uspin}
\end{equation}
are satisfied as far as the relative sign is concerned. 

\subsection{Comparison with other works on \CP violation}
\CP violation in the charmed baryon decays has been recently elaborated on in the final-state rescattering mechanism which amounts to considering the hadronic triangle diagrams in Refs.~\cite{He:2024pxh,Jia:2024pyb}. 
In this subsection we shall compare our work with them.

\subsubsection{Comparison with Jia {\it et al.} \cite{Jia:2024pyb}}
In Ref.~\cite{Jia:2024pyb}, the loop integrals in the hadronic triangle diagrams  are evaluated completely to take into account both real and imaginary parts of amplitudes so that the nontrivial strong phases, which are crucial for computing \CP violating effects, are obtained.  Specifically, branching fractions and \CP asymmetries in the two-body decays $\Lambda_c^+\to \B_8 V$ with a vector meson in the final state are the main focus of   Ref.~\cite{Jia:2024pyb}. 

In this work we also consider the triangle diagrams in Fig. (\ref{Fig:FSI_baryon}). However, we differ from the work of \cite{Jia:2024pyb} in several respects:

\renewcommand{\labelenumi}{\alph{enumi}.}
\begin{enumerate}

\item In our work we did not evaluate the triangle diagrams explicitly. As explained in \cite{He:2024pxh}, our goal is to project out the triangle diagram contributions to penguin and non-penguin amplitudes. {\it A priori} we do not
know how to evaluate the diagrams 
depicted in Fig. \ref{Fig:FSI_baryon} at the quark level which have the penguin topology. At the hadron level, they are manifested as the triangle diagrams and the $s$-channel bubble diagram. However, these diagrams 
consist of not only the penguin topology but also other tree topologies. Therefore, we focus on the flavor 
structure of the triangle and bubble diagrams and project out their contributions to penguin and tree topologies. 

\item Since the loop amplitude of the triangle diagram is divergent and the exchanged particles in the $t$ or $u$ channel are off-shell,  it is necessary to introduce form factors or cutoffs to the strong vertices to render the calculation meaningful in perturbation theory. For example, the form factor is parameterized as \cite{Cheng:FSI}
\be
\label{eq:cutoff}
F=\left( \Lambda^2 \pm m^2\over \Lambda^2\pm p^2\right)^n, \qquad \Lambda =m+\eta\Lambda_{\rm QCD},
\en
where $\pm$ refer to the $s$-channel and $t/u$ channel, respectively. Even though the strong couplings are large in magnitude, the rescattering amplitude is suppressed by a factor of $F^2\sim m^2\Lambda^2_{\rm QCD}/p^4$. Consequently, the off-shell effect will render the perturbative calculations meaningful. 

In our treatment we do not need to specify the $q^2$ dependence of the couplings explicitly as we are concerned with the relation between various parameters.  In other words, we are interested in the decomposition of the triangle or bubble diagram into various IRA or TDA amplitudes, rather than the explicitly detailed expressions of the diagrams. We have assigned overall unknown parameters  $S^-$, $T^-$ and $U^-$, respectively, to $s$-, $t$- and $u$-channel rescattering processes, they will take into account the
 cutoff mechanism.

\item The $s$-channel contributions from Fig. \ref{Fig:FSI_baryon}(a) were neglected in \cite{Jia:2024pyb} based on the argument that the excited states have large width which cannot be neglected
\be
S\sim {1\over p^2-m^2+im\Gamma}
\en
so that the $s$-channel contributions for charmed hadron decay will be suppressed by the width effects of resonant states.  In our work, whether the $s$-channel contribution is negligible or not can be inferred from the overall unknown parameter $S^-$.  From Table \ref{tab:TDAampCoeff} we see that $|T^-|> |S^-|> |U^-|$ from a global fit to the data. Hence, the $s$-channel contribution is not negligible.

\end{enumerate}

\subsubsection{Comparison with He and Liu  \cite{He:2024pxh}}

The final-state rescattering (FSR) approach to charmed baryon CP violation has also been explored recently in Ref. \cite{He:2024pxh}.
The  interaction at the hadron level is obtained by contracting all possible $\text{SU(3)}_F$ indices of antitriplet
charmed baryons, the baryon octet, the meson nonet as well as the $H$ matrices, in analogue  to the 
methodology adopted in the current work. However, 
our work differs from \cite{He:2024pxh} in the following respects:

\renewcommand{\labelenumi}{\roman{enumi}.}

\begin{enumerate}
   
    \item We have included the $u$-channel contribution of FSR, which was neglected in~\cite{He:2024pxh}. As shown explicitly in Eq.~\eqref{eq:biFSI}, the $u$-channel contribution, represented by the unknown parameter $U^-$, appears in all four sets of penguin topology parameters $\tilde b_{1,\ldots,4}$. Whether $U^-$ is negligible or not can be assessed through a global fit. Indeed, we found that $|T^-|> |S^-|> |U^-|$. Recall that  $U^-$ is often accompanied by $\rho_+$, whereas $T^-$ and $S^-$ by $r_\pm$. Since $\rho_+$ is much larger than $r_\pm$ in magnitude, the smallness of $U^-$ is compensated by the largeness of $\rho_+$. Hence, although $|U^-|$ is smaller than $|T^-|$ and $ |S^-|$, the $u$-channel contribution is not negligible.

    \item We have incorporated the hairpin  contribution from the $\tilde b_1$ term in Eq.~\eqref{eq:TDAlambdab2} (or equivalently the $\tilde f^a_3$ term in Eq.~\eqref{eq:IRAHe}), which was not discussed in~\cite{He:2024pxh}. Numerical estimates of the various coefficients, including $\tilde b_i$, are presented in Table~\ref{tab:TDAampCoeff}. The magnitude of $\tilde b_1$ is found to be sizable and should therefore make a significant contribution to the processes involving the SU(3)-singlet $\eta_1$. Moreover, comparing the coefficients $\tilde b_i$ in Table \ref{tab:CompCoeff} with those reported in Ref.~\cite{He:2024pxh}, we find that the sizes of $\tilde b_3$ in $S$-wave and $\tilde b_4$ in $S$- ($P$-)wave are smaller (larger) than those in \cite{He:2024pxh}.
    
    \item The predicted modes with large \CP violation in our analysis differ from those presented in Ref.~\cite{He:2024pxh}.
In particular, we do not observe significant \CPV in the decays $\Xi_c^0 \to \Sigma^+ \pi^-$ and $\Xi_c^0 \to p K^-$.
For example, the \CP asymmetry of $\Xi_c^0 \to \Sigma^+ \pi^-$ was estimated to be 
\(A_{\text{CP}}(\Xi_c^0 \to \Sigma^+ \pi^-) = (0.71 \pm 0.26) \times 10^{-3}\) in~\cite{He:2024pxh}, 
whereas our prediction yields a smaller value of \((0.16 \pm 0.11) \times 10^{-3}\). 
This discrepancy is understandable, as its decay amplitude receives contributions from $\tilde b_2$ and $\tilde b_3$ (see Table~\ref{tab:TDAamp}) and
both of them are subject to the destructive interference between $T^-$ and $U^-$. 
Conversely, our analysis predicts sizable \CP violation in decay modes involving $\eta$ or $\eta'$, as shown in Eq.~(\ref{eq:CPVnum}), 
which were not highlighted in Ref.~\cite{He:2024pxh}. 
This is ascribed to the effect of the hairpin $\tilde b_1$ term in our approach, and the predictions may be tested in the future.

\end{enumerate}

\begin{table}
\caption{ A comparison of coefficients associated with penguin diagrams,
in which the upper entries are results obtained in this work while the lower
entries are quoted from Ref.~
 \cite{He:2024pxh}. Note that the $\tilde f^{b,c,d}_3$ and $ \tilde f^e$ in \cite{He:2024pxh}
 correspond to $\tilde b_{2,3,4}$ and $\tilde b_5$, respectively, in this work.
} 
\label{tab:CompCoeff}
\vspace{-0.1cm}
\begin{center}
\renewcommand\arraystretch{1}
\begin{tabular}
{c| r r r r }
\hline \hline
&$|X_i|_S$&$|X_i|_P$~~~ &$\delta^{X_i}_{S}$~~~~~ &$\delta^{X_i}_{P}$~~~~ \\
&\multicolumn{2}{c}{~~~$(10^{-2}G_{F}~{\rm GeV}^2)$}&\multicolumn{2}{c}{$(\text{in radian})$} \\
\hline
\multirow{2}{*}{$\Tilde{b}_1$}&
~$4.20\pm0.29$&$16.86\pm1.30$&
$2.64\pm0.08$&$-2.03\pm0.08$\\
&$-~~~~~~~$&$-~~~~~~~$&
$-~~~~~~~$&$-~~~~~~~$\\
\multirow{2}{*}{$\Tilde{b}_2$}&
~$0.64\pm0.05$&$10.22\pm0.17$&
$1.09\pm0.08$&$-1.10\pm0.01$\\
&$0.70\pm0.11$&$9.09\pm4.67$&
$0.90\pm0.63$&$-1.01\pm0.07$\\
\multirow{2}{*}{$\Tilde{b}_3$}&
~$1.48\pm0.11$&$11.51\pm0.57$&
$3.04\pm0.10$&$2.50\pm0.05$\\
&$4.37\pm0.99$&$10.46\pm0.65$&
$-3.07\pm0.05$&$-0.69\pm0.13$\\
\multirow{2}{*}{$\Tilde{b}_4$}&
~$3.06\pm0.16$&$11.69\pm0.48$&
$0.23\pm0.04$&$-0.88\pm0.04$\\
&$5.12\pm2.66$&$1.99\pm5.23$&
$0.22\pm0.17$&$-2.07\pm1.32$\\
\multirow{2}{*}{$\Tilde{b}_5$}&
~$1.25\pm0.18$&$0.93\pm0.34$&
$0.30\pm0.11$&$2.76\pm1.35$\\
&$1.17\pm0.74$&$0.40\pm0.63$&
$-2.80\pm0.27$&$-0.97\pm6.10$\\
\hline \hline
\end{tabular}
\end{center}
\end{table}

\subsection{Numerical results for branching fractions, Lee-Yang parameters and phase shifts}

Since there are six more data after our latest global fit \cite{Cheng:2024lsn}, we will present the new fitting
results of branching fractions, Lee-Yang parameters, magnitudes of $S$- and $P$-wave amplitudes and their phase shifts in the TDA  and IRA in Tables \ref{tab:resultScenI}-\ref{tab:fitother2ScenI} in Appendix C.
As noticed in passing, we consider two different fits: Case I without the recent Belle data on $\Xi_c^0\to \Xi^0 \pi^0, \Xi^o\eta^{(')}$  \cite{Belle-II:2024jql} and Case II with all the data included. 
The $\chi^2$ values of the TDA and IRA fits in Cases I and II are shown in Table \ref{tab:chi}. It is clear that the $\chi^2$ value per degree of freedom is around 3.3 in Case I, but slightly higher in Case II.

\begin{table}[t]
\caption{The $\chi^2$ values of the TDA and IRA fits in Cases I and II. }
 \label{tab:chi}
\begin{center} 
\begin{tabular}{l c c c  c}
\hline \hline
& ~~~~Case I~ & & ~~~~Case II & \\
\hline
& TDA & IRA & TDA & IRA \\
\hline
$\chi^2$ & $65.47$ & $65.47$ & $89.53$ & $90.08$ \\
$\chi^2/d.o.f.$ & $3.27$ & $3.27$ & $3.73$ & $3.75$ \\
\hline \hline
\end{tabular}
\end{center} 
\end{table}

In Table~\ref{tab:TDAampCoeff}, we list the numerical values of the coefficients for the tree-level diagrams (also presented in Table II of Ref.~\cite{Cheng:2024lsn}) as well as those for the penguin topological diagrams.

A quantity $\Delta={\rm arg}(H_+/H_-)$, which is the phase difference between the  two helicity amplitudes, was recently measured by LHCb \cite{LHCb:2024tnq}. The decay parameters are then determined by
\begin{equation}
\beta=\sqrt{1-\alpha^2}\,\sin\Delta, \qquad \gamma=\sqrt{1-\alpha^2}\,\cos\Delta.
\end{equation}
We follow our previous analysis \cite{Cheng:2024lsn} to use the new LHCb measurements  to fix the sign ambiguity of $\beta$ and $\gamma$.  As for the phase shift between $S$- and $P$-wave amplitudes, it is naively expected that  $\delta_P - \delta_S = \arctan({\beta}/{\alpha})$ (see, for example, Ref. \cite{BESIII:XiK}). However, this relation often leads to an ambiguity of $\pm\pi$ rad. We have urged to employ a more suitable one \cite{Zhong:2024qqs}
\begin{equation}
\label{eq:phase}
\delta_P - \delta_S = 2 \arctan \frac{\beta}{\sqrt{\alpha^2+\beta^2}+\alpha}.
\end{equation}
It naturally covers the correct solution space without imposing manual modification. 
Likewise, the phase difference $\Delta$ between the two helicity amplitudes is related to the decay
parameters $\beta$ and $\gamma$ by
\begin{equation}
\Delta = 2 \arctan \frac{\beta}{\sqrt{\beta^2+\gamma^2}+\gamma}
\label{eq:Delta}
\end{equation}
in analog to Eq. (\ref{eq:phase}).
The fit results for the phase shift $\delta_P-\delta_S$ and the magnitudes of $S$- and $P$-wave amplitudes are substantially improved over our previous analyses with smaller errors. 

For the decay $\Lambda_c^+\to \Xi^0 K^+$, BESIII found \cite{BESIII:XiK}
\begin{equation}
\alpha=0.01\pm0.16\pm0.03, \quad \beta=-0.64\pm0.69\pm0.13, \quad \gamma=-0.77\pm0.58\pm0.11,
\end{equation}
and uncovered two sets of solutions for the magnitudes of $S$- and $P$-wave amplitudes as well as two solutions for the phase shift, 
\begin{equation}
\label{eq:phaseshifts}
\delta_P-\delta_S=-1.55\pm 0.25\pm0.05~~{\rm or}~~1.59\pm0.25\pm0.05~{\rm rad}.
\end{equation}
Our fits with $\alpha_{\Xi^0K^+}=-0.07\pm0.12$, $\beta_{\Xi^0K^+}=-0.99\pm0.01$ and 
$\delta_P-\delta_S=-1.64\pm 0.12$ rad in Case I are consistent with the first phase-shift solution as well as the Lee-Yang parameters $\alpha_{\Xi^0K^+}$ and $\beta_{\Xi^0K^+}$. As noticed in  Ref.~\cite{Cheng:2024lsn}, 
we have checked that if we set $\delta_S^{X_i}=\delta_P^{X_i}=0$ from the outset and remove the input of $(\alpha_{\Xi^0K^+})_{\rm exp}$ , the fit $\alpha_{\Xi^0K^+}$ will be of order 0.95\,. Hence, we conclude that it is necessary to incorporate the phase shifts to accommodate the data. It is the smallness of $|\cos(\delta_P-\delta_S)|\sim 0.04$ that accounts for the nearly vanishing  $\alpha_{\Xi^0K^+}$.

Besides the decay $\Lambda_c^+\to \Xi^0 K^+$, we have noticed  that  the following modes
$\Xi_c^0\to \Sigma^+ K^-, \Sigma^+\pi^-, p K^-, p\pi^-, n\pi^0$ and $\Xi_c^+\to p \pi^0, n\pi^+$ also  have $\delta_P-\delta_S$ similar to that in $\Lambda_c^+\to \Xi^0 K^+$  and their decay asymmetries will become smaller (see Tables \ref{tab:resultScenI}-\ref{tab:fitother2ScenI}.).
In particular, for the CF channel $\Xi_c^0\to \Sigma^+ K^-$ whose branching fraction has been measured before, 
we predict that $\alpha_{\Xi_c^0\to \Sigma^+ K^-}=-0.07\pm0.12$ in Case I very similar to that of $\Lambda_c^+\to \Xi^0 K^+$. This can be used to test our theoretical framework.

\section{Conclusions}
\label{sec:conclu}
There exist two distinct ways in realizing the approximate SU(3) flavor symmetry to describe the
two-body nonleptonic decays of charmed baryons: the irreducible SU(3) approach (IRA) and the topological diagram approach (TDA). They provide a powerful tool for model-independent analyses. Among them, the IRA has become very popular in recent years.  Nevertheless, the TDA has the advantage that it is more
intuitive, graphic, and easier to implement model calculations. 
The original expression of TDA amplitudes, with 6 more
penguin diagrams included, 
is given in Eq. \eqref{Eq:TDAamp}. 
Five independent TDA tree amplitudes $\tilde T$, $\tilde C$, $\tilde C'$,   $\tilde E_h$ and $\tilde E_1$ are
shown in Eq.  \eqref{eq:tildeTDA}. In terms of the charmed baryon state
$(\B_{c})_i$ and the octet baryon $(\B_8)^i_j$, TDA amplitudes can be
decomposed into $\A_{\rm{TDA}}^{\rm{tree}}$ [see Eq. \eqref{eq:TDAtree2}] 
and $\A_{\rm{TDA}}^{\lambda_b}$ [Eq. \eqref{eq:TDAlambdab2}], 
where the latter is proportional to $\lambda_b$ and has 4 independent amplitudes. 
In the final-state rescattering (FSR) mechanism, sizable penguin topology can be induced 
from LD rescattering. This allows us to establish the relations
between the TDA coefficients in $\A_{\rm{TDA}}^{\lambda_b}$ and $\A_{\rm{TDA}}^{\rm{tree}}$. 
Based on the TDA at the tree level, we have updated the global fits to the currently available data of two-body charmed baryon decays, 
including the recent 6 measurements of $\Xi_c^+$ decays
performed by Belle and Belle-II.

Phenomenological implications are
as follows:
\renewcommand{\labelenumi}{\arabic{enumi}.}

\begin{enumerate}
\item 
The equivalence of the TDA and IRA is formally demonstrated in Sec.II.B. The numerical analyses within their errors do support the conclusion that the two approaches are equivalent, although some differences
are observed  in the uncertainties of certain observables,
originated from the inputs of fitted parameters that propagated and pronounced by nonlinear derivatives.
 
\item 
We urge to apply Eqs. (\ref{eq:phase}) and (\ref{eq:Delta}) to extract the phase shift $\delta_P-\delta_S$
from the decay parameters $\alpha$ and $\beta$
and the phase difference $\Delta$ between the two helicity amplitudes from $\beta$ and $\gamma$, respectively, to avoid a possible ambiguity of $\pm \pi$.

\item 
Since we use the new LHCb measurements to fix the sign ambiguity of $\beta$ and $\gamma$, the fit results for the phase shift $\delta_P-\delta_S$ and the magnitudes of $S$- and $P$-wave amplitudes are substantially improved over our previous analyses with smaller errors. The fitting 
branching fractions, decay parameters, magnitudes of $S$- and $P$-wave amplitudes and their phase shifts in both the TDA and IRA are presented in Tables \ref{tab:resultScenI}-\ref{tab:fitother2ScenI}. These results can be tested in the near future.

\item Final-state $s$-, $t$- and $u$-channel rescattering contributions to $\B_c\to P\B$ from the external $W$-emission are depicted in Figs. \ref{Fig:FSI_baryon}(a), \ref{Fig:FSI_baryon}(b) and \ref{Fig:FSI_baryon}(c), respectively.  They have the same topology as the penguin one.
Although we do not know how to evaluate these diagrams at the quark level, their contributions at the hadron level manifested as the pole and triangle diagrams are amenable. We have followed the work of He and Liu \cite{He:2024pxh} to assign the FSR (final-state rescattering) parameters $S^-$, $T^-$ and $U^-$ as the overall unknown parameters to the respective channels.  The FSR parameters and  TDA penguin coefficients then can be resolved
via Eqs.  ~\eqref{eq:TDAtree} and ~\eqref{eq:biFSI}.
We found that $|T^-|> |S^-|> |U^-|$ for both $S$- and $P$-wave transitions from a global fit to the data. Hence, the $u$-channel contribution is small compared to $t$- and $s$-channel ones, nevertheless it is not entirely negligible.   Recall that the parameter $\rho_+$ in the $u$-channel is much larger in magnitude than the parameters $r_\pm$ in $s$- and $t$-channel. Moreover, it is not justified to neglect the $s$-channel contribution as claimed  in Ref.~\cite{Jia:2024pyb}.

\item For the $u$-channel decay amplitude of $\B_c\to P\B$ mediated by the vector-meson exchange, we have made the assumption that it  is dominated by the tensor coupling $\tilde f_{V\B'\B}$.

\item The predicted {\it CP}-violating observables are summarized in Table~\ref{tab:CPobsv}. 
\CP asymmetries at the per mille are found in the following decay modes:
\begin{equation*}
\Lambda_c \to p \pi^0,\qquad 
\Lambda_c \to p \eta',\qquad 
\Xi_c^0 \to \Sigma^0 \eta,\qquad 
\Xi_c^+ \to \Sigma^+ \eta.
\end{equation*}
In particular, we do not observe significant \CP violation in the decays $\Xi_c^0 \to \Sigma^+ \pi^-$ and $\Xi_c^0 \to p K^-$ as advocated in Ref.~\cite{He:2024pxh}.
This discrepancy is traced back to the decay amplitude in our framework which receives contributions from        $\tilde b_2$ and $\tilde b_3$,
both of which are subject to a destructive interference between $T^-$ and $U^-$. 

\end{enumerate}

\section*{Acknowledgments}

We are grateful to  Cai-Ping Jia, Chia-Wei Liu and Fu-Sheng Yu for valuable discussions. We also acknowledge Yixuan Wu for her contribution in cross-checking part of the calculations.
This research was supported in part by the National Science and Technology Council of R.O.C. under Grant No.~112-2112-M-001-026  and the National Natural Science Foundation of China under Grant Nos.~12475095 and U1932104.

\appendix
\section{Strong $P\B\B$ couplings}
The interaction of the pseudoscalar meson with octet baryons under SU(3) symmetry has the general
expression
\be
{\cal L}_{P\B\B}&=& g_1\bar \B i\gamma_5 D^a B P_a+g_2\bar \B i\gamma_5 F^a B P_a \non \\
 &=& g\,\bar \B i\gamma_5 \left[ \alpha D^a +(1-\alpha)F^a\right] B P_a,
 \en
where the $D$- and $F$-type matrices can be found in \cite{Carr}.  Hence, one needs two parameters $g$ and  $\alpha$ to describe the $P\B\B$ couplings. Alternatively, we follow \cite{He:2024pxh} to write
\be
\sum_{\B_+,\B,P} g_{\B_+\B P}P\ov \B i\gamma_5\B_+    &=&
 g_+\left[ (P)^i_j(\ov\B)^j_ki\gamma_5(B_+)^k_i+ r_+(P)^j_k(\ov \B)^i_j i\gamma_5(B_+)^k_i\right],  \non \\
 \sum_{\B_-,\B,P} g_{\B_-\B P}P\ov \B \B_-    &=&
 g_-\left[ (P)^i_j(\ov\B)^j_k(B_-)^k_i+ r_-(P)^j_k(\ov \B)^i_j (B_-)^k_i\right], 
\en
where the subscript $\pm$ denotes the positive (negative) parity of the baryon $\B$.  The $P\B\B_\pm$ couplings are then expressed in terms of $g_\pm$ and $r_\pm$. For example, 
\be 
\label{eq:PBB}
&& g_{n\Sigma^-K^+}=g_+, \quad g_{\Lambda p K^+}={g_+\over\sqrt{6}}(1-2r_+), \non \\
&& g_{\Lambda\Sigma^- \pi^+}={g_+\over\sqrt{6}}(1+ r_+),  \quad
g_{\Sigma^0 \Sigma^+ \pi^-}={g_+\over\sqrt{2}}(1-r_+).
\en
Using the input of $g_{n\Sigma^-K^+}=6.70$ and $g_{\Lambda p K^+}=-10.78$, He and Liu \cite{He:2024pxh} obtained $g_+=6.70$ and $r_+=2.47$. However, if the input of $g_{n\Sigma^-K^+}=4.95$,       
   $g_{\Lambda p K^+}=-15.2$, $g_{\Lambda\Sigma^- \pi^+}=10.6$ and $g_{\Sigma^0 \Sigma^+ \pi^-}=-13.3$ 
from \cite{Verma:PBB} is employed, we will have $g_+=4.95$ and $r_+=4.26$.

\section{Experimental data}
\label{app:exp}

The latest experimental data on branching fractions and Lee-Yang parameters 
of antitriplet charmed baryon decays are summarized in Table \ref{tab:expdata}.

\thispagestyle{empty} 
\begin{table}[tp!]
\caption{
Updated experimental data on branching fractions and Lee-Yang parameters.
The upper entries correspond to those used in the earlier analysis \cite{Cheng:2024lsn},
while the lower entries include the latest measurements from Belle and Belle II 
~\cite{Belle:2024xcs,Belle-II:2025klu}.
} 
\vspace{-0.4cm}
\label{tab:expdata}
\thispagestyle{empty} 
\begin{center}
\renewcommand\arraystretch{1}
\resizebox{\textwidth}{!} 
{
\begin{tabular}
{ l c c c c c }
\hline
Observable & PDG \cite{PDG} & BESIII & Belle & LHCb & Average\\
\hline
$10^{2}\mathcal{B}(\Lambda_c^+\to\Lambda^0\pi^+)$&
$1.29\pm{0.05}$&&& & $1.29\pm{0.05}$\\

$10^{2}\mathcal{B}(\Lambda_c^+\to \Sigma^0 \pi^+)$&
$1.27\pm{0.06}$&&& & 
$1.27\pm{0.06}$\\

$10^{2}\mathcal{B}(\Lambda_c^+\to \Sigma^+ \pi^0)$&
$1.24\pm0.09$&&& &
$1.24\pm0.09$\\

$10^{2}\mathcal{B}(\Lambda_c^+\to \Sigma^+ \eta)$&
$0.32\pm0.05$&&& & 
$0.32\pm0.05$\\

$10^{2}\mathcal{B}(\Lambda_c^+\to \Sigma^+ \eta')$&
$0.41\pm0.08$&&& &
$0.41\pm0.08$\\

$10^{2}\mathcal{B}(\Lambda_c^+\to \Xi^0 K^+)$&
$0.55\pm0.07$&&&  & 
$0.55\pm0.07$\\

$10^{4}\mathcal{B}(\Lambda_c^+\to \Lambda^0 K^+)$&
$6.42\pm0.31$& & & &
$6.42\pm0.31$\\

$10^{4}\mathcal{B}(\Lambda_c^+\to \Sigma^0 K^+)$&
$3.70\pm0.31$& & & &
$3.70\pm0.31$\\

$10^{4}\mathcal{B}(\Lambda_c^+\to \Sigma^+ K_S)$&
$4.7\pm1.4$&& & &
$4.7\pm1.4$\\

$10^{4}\mathcal{B}(\Lambda_c^+\to n \pi^+)$&
$6.6\pm1.3$&&& &
$6.6\pm1.3$\\

$10^{4}\mathcal{B}(\Lambda_c^+\to p\pi^0)$&
$<0.8$& $1.56^{+0.75}_{-0.61}$ & &  &
$1.56^{+0.75}_{-0.61} $\\

$10^{2}\mathcal{B}(\Lambda_c^+\to p K_S)$&
$1.59\pm0.07$&&& &
$1.59\pm0.07$\\

$10^{3}\mathcal{B}(\Lambda_c^+\to p \eta)$&
$1.57\pm0.12$ &  $1.63\pm0.33$ & & &$1.58\pm0.11$\\

$10^{4}\mathcal{B}(\Lambda_c^+\to p \eta')$&
$4.8\pm0.9$&&& &
$4.8\pm0.9$\\

$10^{2}\mathcal{B}(\Xi_c^0\to \Xi^- \pi^+)$&
$1.43\pm0.27$&& 
$1.80\pm0.52$& &
$1.80\pm0.52$\\

$10^{2}\frac{\mathcal{B}(\Xi_c^0\to \Xi^- K^+)}{\mathcal{B}(\Xi_c^0\to \Xi^- \pi^+)}$&
$2.75\pm0.57$&&&& $2.75\pm0.57$\\

$10^{2}\frac{\mathcal{B}(\Xi_c^0\to \Lambda K_S^0)}{\mathcal{B}(\Xi_c^0\to \Xi^- \pi^+)}$&
$22.5\pm1.3$&&& & $22.5\pm1.3$\\

$10^{2}\frac{\mathcal{B}(\Xi_c^0\to \Sigma^0 K_S^0)}{\mathcal{B}(\Xi_c^0\to \Xi^- \pi^+)}$&
$3.8\pm0.7$&&& & $3.8\pm0.7$\\

$10^{2}\frac{\mathcal{B}(\Xi_c^0\to \Sigma^+ K^-)}{\mathcal{B}(\Xi_c^0\to \Xi^- \pi^+)}$&
$12.3\pm1.2$&&& & $12.3\pm1.2$\\

$10^{3}\mathcal{B}(\Xi_c^0\to\Xi^0\pi^0)$  &  & & $6.9\pm1.6$ & & $6.9\pm1.6$\\

$10^{3}\mathcal{B}(\Xi_c^0\to\Xi^0\eta)$ & & & $1.6\pm0.5$ & & $1.6\pm0.5$\\

$10^{3}\mathcal{B}(\Xi_c^0\to\Xi^0\eta')$  & & & $1.2\pm0.4$ & & $1.2\pm0.4$ \\

$10^{2}\mathcal{B}(\Xi_c^+\to \Xi^0 \pi^+)$&
$1.6\pm0.8$&& {$0.719\pm0.323$~\cite{Belle-II:2025klu}}& & {$0.84\pm0.48$} \\

$\alpha(\Lambda_c^+\to \Lambda^0 \pi^+)$& 
$-0.755\pm0.006$ & & & $-0.782\pm 0.010$ & {$-0.762\pm0.006$} \\

$\alpha(\Lambda_c^+\to \Sigma^0 \pi^+)$&
$-0.466\pm0.018$ & & & & $-0.466\pm0.018$ \\

$\alpha(\Lambda_c^+\to p K_S)$&
{$0.18\pm0.45$}&&& $-0.744\pm0.015$ & {$-0.743\pm0.028$} \\

$\alpha(\Lambda_c^+\to \Sigma^+ \pi^0)$&
$-0.484\pm0.027$ & & & & $-0.484\pm0.027$ \\

$\alpha(\Lambda_c^+\to \Sigma^+ \eta)$&
$-0.99\pm0.06$ & & & & $-0.99\pm0.06$\\

$\alpha(\Lambda_c^+\to \Sigma^+ \eta')$&
$-0.46\pm0.07$ & & & & $-0.46\pm0.07$\\

$\alpha(\Lambda_c^+\to \Lambda^0 K^+)$&
{$-0.585\pm0.052$} & & & $-0.569\pm0.065$ &{$-0.579\pm0.041$}\\

$\alpha(\Lambda_c^+\to \Sigma^0 K^+)$&
$-0.54\pm0.20$ & & & & $-0.54\pm0.20$\\

$\alpha(\Lambda_c^+\to \Xi^0 K^+)$&
&$0.01\pm0.16$ && & $0.01\pm0.16$\\

$\alpha(\Xi_c^0\to \Xi^- \pi^+)$&
$-0.64\pm0.05$&&&& $-0.64\pm0.05$\\

$\alpha(\Xi_c^0\to\Xi^0\pi^0)$ & & & $-0.90\pm0.27$ & & $-0.90\pm0.27$\\

$\beta(\Lambda_c^+\to \Lambda^0 \pi^+)$& 
& & & $0.368\pm0.021$& $0.368\pm0.021$ \\
$\beta(\Lambda_c^+\to \Lambda^0 K^+)$& 
& & & $0.35\pm0.13$ &  $0.35\pm0.13$\\

$\gamma(\Lambda_c^+\to \Lambda^0 \pi^+)$& 
& & & $0.502\pm0.017$ & $0.502\pm0.017$  \\
$\gamma(\Lambda_c^+\to \Lambda^0 K^+)$& 
& & & $-0.743\pm0.071$ & $-0.743\pm0.071$ \\

\hline

{$10^{2}\mathcal{B}(\Xi_c^+\to \Sigma^+ K_S^0)$}&
&&$0.194\pm0.090$~\cite{Belle-II:2025klu}& & $0.194\pm0.090$ \\

{$10^{2}\mathcal{B}(\Xi_c^+\to \Xi^0 \pi^+)$}&
&&$0.719\pm0.323$~\cite{Belle-II:2025klu}& & $0.719\pm0.323$ \\

{$10^{2}\mathcal{B}(\Xi_c^+\to \Xi^0 K^+)$}&
&&$0.049\pm0.023$~\cite{Belle-II:2025klu}& & $0.049\pm0.023$ \\

{$10^{4}\mathcal{B}(\Xi_c^+\to p K_S^0)$}&
&&$7.16\pm3.25$~\cite{Belle:2024xcs}& &$7.16\pm3.25$\\

{$10^{4}\mathcal{B}(\Xi_c^+\to \Lambda \pi^+)$}&
&&$4.52\pm2.09$~\cite{Belle:2024xcs}& &$4.52\pm2.09$\\

{$10^{4}\mathcal{B}(\Xi_c^+\to \Sigma^0 \pi^+)$}&
&&$1.20\pm0.55$~\cite{Belle:2024xcs}& &$1.20\pm0.55$\\

\hline
\end{tabular}
}
\end{center}
\end{table}
\thispagestyle{empty}

\section{New fitting
results of branching fractions, Lee-Yang parameters and etc.}
\label{app:fit}

The new fitting
results of branching fractions, Lee-Yang parameters, magnitudes of $S$- and $P$-wave amplitudes and their phase shifts in the TDA  and IRA are presented in Tables \ref{tab:resultScenI}-\ref{tab:fitother2ScenI}.

\begin{table*}[h]
\caption{
The fit results based on the tilde TDA (upper) and tilde IRA (lower) in Case I. Experimental data from Belle \cite{Belle-II:2024jql}  with an asterisk are not included in the fit.
$S$- and $P$-wave amplitudes are in units of $10^{-2}G_F~{\rm GeV}^2$ and $\delta_P-\delta_S$ in radian. }
\label{tab:resultScenI}
\centering
\resizebox{\textwidth}{!} 
{
\begin{tabular}
{ l |c rrr rrr |  c c  c }
\hline
\hline
\multirow{2}{*}{Channel}
&\multirow{2}{*}{$10^{2}\mathcal{B}$}
&\multirow{2}{*}{$\alpha$}~~~~~~~
&\multirow{2}{*}{$\beta$}~~~~~~~
&\multirow{2}{*}{$\gamma$}~~~~~~
&\multirow{2}{*}{$|A|$}~~~~
&\multirow{2}{*}{$|B|$}~~~~ 
&\multirow{2}{*}{$\delta_P-\delta_S$}~~~ 
&\multirow{2}{*}{$10^{2}\mathcal{B}_\text{exp}$}
&\multirow{2}{*}{$\alpha_\text{exp}$}
&$\beta_\text{exp}$\\
&&&&&&&&&&$\gamma_\text{exp}$\\
\hline
\multirow{2}{*}{$\Lambda_c^+\to\Lambda^0\pi^+$}&$1.25\pm0.04$&$-0.76\pm0.01$&$0.39\pm0.02$&$0.51\pm0.01$&$5.46\pm0.10$&$9.04\pm0.19$&$2.67\pm0.02$&\multirow{2}{*}{$1.29\pm{0.05}$}&\multirow{2}{*}{$-0.762\pm0.006$}&$0.368\pm0.021$\\
&$1.25\pm0.04$&$-0.76\pm0.01$&$0.39\pm0.02$&$0.51\pm0.01$&$5.46\pm0.10$&$9.04\pm0.19$&$2.67\pm0.02$&&&$0.502\pm0.017$\\
\multirow{2}{*}{$\Lambda_c^+\to\Sigma^0\pi^+$}&$1.21\pm0.05$&$-0.47\pm0.01$&$0.48\pm0.09$&$-0.74\pm0.06$&$2.22\pm0.24$&$18.51\pm0.50$&$2.35\pm0.10$&\multirow{2}{*}{$1.27\pm{0.06}$}&\multirow{2}{*}{$-0.466\pm0.018$}\\
&$1.21\pm0.05$&$-0.47\pm0.01$&$0.48\pm0.09$&$-0.74\pm0.06$&$2.22\pm0.24$&$18.51\pm0.50$&$2.35\pm0.10$\\
\multirow{2}{*}{$\Lambda_c^+\to\Sigma^+ \pi^0$}&$1.22\pm0.05$&$-0.47\pm0.01$&$0.47\pm0.09$&$-0.75\pm0.06$&$2.22\pm0.24$&$18.51\pm0.50$&$2.35\pm0.10$&\multirow{2}{*}{$1.24\pm0.09$}&\multirow{2}{*}{$-0.484\pm0.027$}\\&$1.22\pm0.05$&$-0.47\pm0.01$&$0.47\pm0.09$&$-0.75\pm0.06$&$2.22\pm0.24$&$18.51\pm0.50$&$2.35\pm0.10$\\
\multirow{2}{*}{$\Lambda_c^+\to\Sigma^+ \eta$}&$0.35\pm0.04$&$-0.91\pm0.04$&$-0.08\pm0.13$&$0.42\pm0.09$&$3.03\pm0.21$&$7.02\pm0.68$&$-3.05\pm0.15$&\multirow{2}{*}{$0.32\pm0.05$}&\multirow{2}{*}{$-0.99\pm0.06$}\\&$0.35\pm0.04$&$-0.91\pm0.04$&$-0.08\pm0.13$&$0.42\pm0.10$&$3.03\pm0.21$&$7.02\pm0.68$&$-3.05\pm0.14$\\
\multirow{2}{*}{$\Lambda_c^+\to\Sigma^+ \eta'$}&$0.40\pm0.07$&$-0.43\pm0.07$&$0.87\pm0.07$&$0.22\pm0.24$&$4.22\pm0.71$&$21.01\pm2.55$&$2.03\pm0.08$&\multirow{2}{*}{$0.41\pm0.08$}&\multirow{2}{*}{$-0.46\pm0.07$}\\&$0.40\pm0.07$&$-0.43\pm0.07$&$0.87\pm0.07$&$0.22\pm0.24$&$4.22\pm0.71$&$21.01\pm2.53$&$2.03\pm0.08$\\
\multirow{2}{*}{$\Lambda_c^+\to\Xi^0 K^+$}&$0.33\pm0.03$&$-0.07\pm0.12$&$-0.99\pm0.01$&$0.13\pm0.09$&$2.65\pm0.18$&$9.90\pm0.45$&$-1.64\pm0.12$&\multirow{2}{*}{$0.55\pm0.07$}&\multirow{2}{*}{$0.01\pm0.16$}\\&$0.33\pm0.03$&$-0.07\pm0.12$&$-0.99\pm0.01$&$0.13\pm0.09$&$2.65\pm0.18$&$9.90\pm0.45$&$-1.64\pm0.12$\\
\multirow{2}{*}{$\Lambda_c^+\to p K_S$}&$1.57\pm0.06$&$-0.74\pm0.03$&$0.59\pm0.12$&$-0.32\pm0.21$&$4.36\pm0.66$&$15.43\pm1.28$&$2.47\pm0.10$&\multirow{2}{*}{$1.59\pm0.07$}&\multirow{2}{*}{$-0.743\pm0.028$}\\&$1.57\pm0.06$&$-0.74\pm0.03$&$0.59\pm0.12$&$-0.32\pm0.21$&$4.36\pm0.66$&$15.43\pm1.29$&$2.47\pm0.10$\\
\multirow{2}{*}{$\Xi_c^0\to\Xi^- \pi^+$}&$2.96\pm0.08$&$-0.72\pm0.03$&$0.67\pm0.02$&$0.17\pm0.04$&$8.20\pm0.20$&$23.01\pm0.58$&$2.39\pm0.04$&\multirow{2}{*}{$1.80\pm0.52$}&\multirow{2}{*}{$-0.64\pm0.05$}\\&$2.96\pm0.08$&$-0.72\pm0.03$&$0.67\pm0.03$&$0.17\pm0.04$&$8.20\pm0.20$&$23.01\pm0.58$&$2.39\pm0.04$\\
\multirow{2}{*}{$\Xi_c^0\to\Xi^0 \pi^0$}&$0.73\pm0.04$&$-0.62\pm0.07$&$0.78\pm0.05$&$0.08\pm0.10$&$3.93\pm0.24$&$11.92\pm0.61$&$2.25\pm0.09$&\multirow{2}{*}{$0.69\pm0.16^{*}$}&\multirow{2}{*}{$-0.90\pm0.27^{*}$}\\&$0.73\pm0.04$&$-0.62\pm0.07$&$0.78\pm0.05$&$0.08\pm0.10$&$3.93\pm0.23$&$11.92\pm0.61$&$2.25\pm0.09$\\
\multirow{2}{*}{$\Xi_c^0\to\Xi^0 \eta$}&$0.24\pm0.04$&$0.23\pm0.15$&$0.04\pm0.15$&$-0.97\pm0.03$&$0.38\pm0.24$&$11.91\pm0.98$&$0.19\pm0.68$&\multirow{2}{*}{$0.16\pm0.05^{*}$}\\&$0.24\pm0.04$&$0.23\pm0.15$&$0.04\pm0.15$&$-0.97\pm0.03$&$0.38\pm0.24$&$11.91\pm0.97$&$0.19\pm0.68$\\
\multirow{2}{*}{$\Xi_c^0\to\Xi^0 \eta'$}&$0.44\pm0.06$&$-0.68\pm0.06$&$0.73\pm0.06$&$-0.02\pm0.25$&$4.04\pm0.69$&$23.43\pm2.48$&$2.32\pm0.08$&\multirow{2}{*}{$0.12\pm0.04^{*}$}\\&$0.44\pm0.06$&$-0.68\pm0.06$&$0.73\pm0.06$&$-0.02\pm0.24$&$4.04\pm0.68$&$23.43\pm2.45$&$2.32\pm0.08$\\
\multirow{2}{*}{$\Xi_c^+\to\Xi^0\pi^+$}&$0.85\pm0.10$&$-0.89\pm0.07$&$0.37\pm0.10$&$0.27\pm0.13$&$2.65\pm0.19$&$6.63\pm0.73$&$2.75\pm0.12$&\multirow{2}{*}{$0.84\pm0.48$}\\&$0.85\pm0.10$&$-0.89\pm0.07$&$0.37\pm0.11$&$0.27\pm0.13$&$2.65\pm0.19$&$6.63\pm0.74$&$2.75\pm0.12$\\
\multirow{2}{*}{$\Lambda_c^+\to\Lambda^0 K^+$}&$0.0635\pm0.0029$&$-0.57\pm0.04$&$0.41\pm0.07$&$-0.71\pm0.04$&$0.57\pm0.04$&$4.41\pm0.11$&$2.52\pm0.10$&\multirow{2}{*}{$0.0642\pm0.0031$}&\multirow{2}{*}{$-0.579\pm0.041$}&$0.35\pm0.13$\\
&$0.0635\pm0.0029$&$-0.57\pm0.04$&$0.41\pm0.07$&$-0.71\pm0.04$&$0.57\pm0.04$&$4.41\pm0.11$&$2.52\pm0.10$&&&$-0.743\pm0.071$\\
\multirow{2}{*}{$\Lambda_c^+\to\Sigma^0 K^+$}&$0.0393\pm0.0022$&$-0.62\pm0.07$&$0.77\pm0.05$&$0.15\pm0.10$&$0.90\pm0.05$&$2.75\pm0.14$&$2.25\pm0.09$&\multirow{2}{*}{$0.0370\pm0.0031$}&\multirow{2}{*}{$-0.54\pm0.20$}\\&$0.0393\pm0.0021$&$-0.62\pm0.07$&$0.77\pm0.05$&$0.15\pm0.10$&$0.90\pm0.05$&$2.75\pm0.14$&$2.25\pm0.09$\\
\multirow{2}{*}{$\Lambda_c^+\to\Sigma^+K_S$}&$0.0394\pm0.0022$&$-0.62\pm0.07$&$0.77\pm0.05$&$0.15\pm0.10$&$0.90\pm0.05$&$2.75\pm0.14$&$2.25\pm0.09$&\multirow{2}{*}{$0.047\pm0.014$}&\\&$0.0394\pm0.0021$&$-0.62\pm0.07$&$0.77\pm0.05$&$0.15\pm0.10$&$0.90\pm0.05$&$2.75\pm0.14$&$2.25\pm0.09$\\
\multirow{2}{*}{$\Lambda_c^+\to n\pi^+$}&$0.061\pm0.006$&$-0.51\pm0.11$&$-0.72\pm0.04$&$0.46\pm0.11$&$1.20\pm0.06$&$1.76\pm0.23$&$-2.19\pm0.12$&\multirow{2}{*}{$0.066\pm0.013$}&\\&$0.061\pm0.006$&$-0.51\pm0.12$&$-0.72\pm0.04$&$0.46\pm0.11$&$1.20\pm0.06$&$1.76\pm0.23$&$-2.19\pm0.12$\\
\multirow{2}{*}{$\Lambda_c^+\to p\pi^0$}&$0.0188\pm0.0033$&$-0.46\pm0.36$&$-0.88\pm0.26$&$-0.15\pm0.47$&$0.51\pm0.13$&$1.41\pm0.36$&$-2.05\pm0.45$&\multirow{2}{*}{$0.0156_{-0.0061}^{+0.0075}$}&\\&$0.0188\pm0.0033$&$-0.46\pm0.36$&$-0.88\pm0.26$&$-0.15\pm0.47$&$0.51\pm0.13$&$1.41\pm0.36$&$-2.05\pm0.45$\\
\multirow{2}{*}{$\Lambda_c^+\to p\eta$}&$0.163\pm0.009$&$-0.64\pm0.06$&$0.49\pm0.18$&$-0.59\pm0.17$&$1.11\pm0.24$&$5.60\pm0.31$&$2.49\pm0.18$&\multirow{2}{*}{$0.158\pm0.011$}&\\&$0.163\pm0.009$&$-0.64\pm0.06$&$0.49\pm0.18$&$-0.59\pm0.17$&$1.11\pm0.24$&$5.60\pm0.31$&$2.49\pm0.18$\\
\multirow{2}{*}{$\Lambda_c^+\to p\eta'$}&$0.052\pm0.008$&$-0.46\pm0.10$&$0.69\pm0.17$&$-0.56\pm0.2$&$0.77\pm0.16$&$4.70\pm0.54$&$2.17\pm0.17$&\multirow{2}{*}{$0.048\pm0.009$}&\\&$0.052\pm0.008$&$-0.46\pm0.10$&$0.69\pm0.17$&$-0.56\pm0.2$&$0.77\pm0.16$&$4.70\pm0.54$&$2.17\pm0.17$\\

\multirow{2}{*}{$\Xi_c^{+} \rightarrow p K_{S}$}
&$0.151\pm0.008$&$-0.60\pm0.06$&$0.75\pm0.07$&$-0.26\pm0.09$&$0.90\pm0.05$&$2.75\pm0.14$&$2.25\pm0.09$&{\multirow{2}{*}{$0.0716\pm0.0325$}}&\\
&$0.151\pm0.008$&$-0.60\pm0.06$&$0.75\pm0.07$&$-0.26\pm0.09$&$0.90\pm0.05$&$2.75\pm0.14$&$2.25\pm0.09$\\
\multirow{2}{*}{$\Xi_c^{+} \rightarrow \Lambda^0 \pi^{+}$}&$0.021\pm0.004$&$-0.27\pm0.40$&$0.14\pm0.25$&$0.95\pm0.10$&$0.52\pm0.05$&$0.22\pm0.23$&$2.67\pm1.17$&{\multirow{2}{*}{$0.0452\pm0.0209$}}&\\&$0.021\pm0.004$&$-0.27\pm0.41$&$0.14\pm0.25$&$0.95\pm0.10$&$0.52\pm0.05$&$0.22\pm0.23$&$2.67\pm1.34$\\
\multirow{2}{*}{$\Xi_c^{+} \rightarrow \Sigma^0 \pi^{+}$}&$0.316\pm0.009$&$-0.59\pm0.02$&$0.57\pm0.05$&$-0.57\pm0.04$&$0.94\pm0.05$&$5.19\pm0.10$&$2.38\pm0.06$&{\multirow{2}{*}{$0.120\pm0.055$}}&\\&$0.316\pm0.009$&$-0.59\pm0.02$&$0.57\pm0.05$&$-0.57\pm0.04$&$0.94\pm0.05$&$5.19\pm0.10$&$2.38\pm0.06$\\
\multirow{2}{*}{$\Xi_c^{+} \rightarrow \Sigma^{+} K_S$}
&$0.191\pm0.065$&$-0.76\pm0.67$&$-0.07\pm0.38$&$-0.65\pm0.76$&$0.69\pm0.67$&$4.64\pm1.74$&$-3.05\pm0.54$&{\multirow{2}{*}{$0.194\pm0.090$}}&\\&$0.191\pm0.065$&$-0.76\pm0.67$&$-0.07\pm0.38$&$-0.65\pm0.76$&$0.69\pm0.67$&$4.64\pm1.73$&$-3.05\pm0.54$\\
{\multirow{2}{*}{$\Xi_c^{+} \rightarrow \Xi^0 K^{+}$}}&$0.114\pm0.010$&$-0.40\pm0.1$&$-0.57\pm0.03$&$0.72\pm0.07$&$1.20\pm0.06$&$1.76\pm0.23$&$-2.19\pm0.12$&{\multirow{2}{*}{$0.049\pm0.023$}}&\\&$0.114\pm0.010$&$-0.40\pm0.1$&$-0.57\pm0.03$&$0.72\pm0.07$&$1.20\pm0.06$&$1.76\pm0.23$&$-2.19\pm0.12$\\
\hline
Channel
&$10^{2}\mathcal{R}_X$&$\alpha$~~~~~~~&$\beta$~~~~~~~&$\gamma$~~~~~~
&$|A|$~~~~&$|B|$~~~~ &$\delta_P-\delta_S$~~~ 
&$10^{2}(\mathcal{R}_X)_\text{exp}$
&$\alpha_\text{exp}$
&$\beta_\text{exp}$
\\
\hline
\multirow{2}{*}{$\Xi_c^0\to\Lambda K_S^0$}&$22.79\pm0.76$&$-0.61\pm0.02$&$0.60\pm0.09$&$-0.51\pm0.12$&$2.66\pm0.32$&$13.15\pm0.52$&$2.36\pm0.08$&\multirow{2}{*}{$22.9\pm1.3$}&\\&$22.79\pm0.76$&$-0.61\pm0.02$&$0.60\pm0.09$&$-0.51\pm0.12$&$2.66\pm0.32$&$13.15\pm0.52$&$2.36\pm0.08$\\
\multirow{2}{*}{$\Xi_c^0\to\Sigma^0 K_S^0$}&$3.76\pm0.63$&$-0.54\pm0.41$&$-0.81\pm0.14$&$0.24\pm0.50$&$1.72\pm0.41$&$4.15\pm1.30$&$-2.16\pm0.43$&\multirow{2}{*}{$3.8\pm0.7$}&\\&$3.76\pm0.63$&$-0.54\pm0.41$&$-0.81\pm0.14$&$0.24\pm0.50$&$1.72\pm0.41$&$4.15\pm1.30$&$-2.16\pm0.43$\\
\multirow{2}{*}{$\Xi_c^0\to\Sigma^+K^-$}&$13.76\pm0.94$&$-0.07\pm0.12$&$-0.98\pm0.02$&$-0.19\pm0.09$&$2.65\pm0.18$&$9.90\pm0.45$&$-1.64\pm0.12$&\multirow{2}{*}{$12.3\pm1.2$}&\\&$13.76\pm0.94$&$-0.07\pm0.12$&$-0.98\pm0.02$&$-0.19\pm0.09$&$2.65\pm0.18$&$9.90\pm0.45$&$-1.64\pm0.12$\\
\multirow{2}{*}{$\Xi_c^0\to\Xi^-K^+$}&$4.40\pm0.02$&$-0.71\pm0.03$&$0.66\pm0.02$&$0.25\pm0.04$&$1.89\pm0.05$&$5.30\pm0.13$&$2.39\pm0.04$&\multirow{2}{*}{$2.75\pm0.57$}&\\&$4.40\pm0.02$&$-0.71\pm0.03$&$0.66\pm0.02$&$0.25\pm0.04$&$1.89\pm0.05$&$5.30\pm0.13$&$2.39\pm0.04$\\
\hline
\hline
\end{tabular}
}
\end{table*}

\begin{table*}[tp!]\footnotesize
\caption{
Same as Table \ref{tab:resultScenI} except for yet-observed CF and SCS  modes in Case I.
}
\label{tab:fitotherScenI}
 \resizebox{\textwidth}{!} 
 {
\centering
\begin{tabular}
{ l | r rrr rr r
}
\hline
\hline
Channel&
$10^{3}\mathcal{B}$~~~~ &$\alpha$~~~~~~~&$\beta$~~~~~~~&$\gamma$~~~~~~
&$|A|$~~~ & $|B|$~~~ & $\delta_P-\delta_S$~~~\\
\hline
\multirow{2}{*}{$\Lambda_c^{+} \rightarrow p K_L$}&$15.09\pm0.54$&$-0.73\pm0.03$&$0.60\pm0.11$&$-0.31\pm0.19$&$4.31\pm0.60$&$15.06\pm1.16$&$2.45\pm0.10$\\&$15.09\pm0.54$&$-0.73\pm0.03$&$0.60\pm0.11$&$-0.31\pm0.19$&$4.31\pm0.60$&$15.06\pm1.17$&$2.45\pm0.10$\\
\multirow{2}{*}{$\Xi_c^0 \rightarrow \Lambda^0 K_L$}&$7.50\pm0.21$&$-0.59\pm0.02$&$0.59\pm0.09$&$-0.56\pm0.10$&$2.66\pm0.31$&$14.09\pm0.50$&$2.35\pm0.07$\\&$7.50\pm0.21$&$-0.59\pm0.02$&$0.59\pm0.09$&$-0.56\pm0.10$&$2.66\pm0.31$&$14.09\pm0.50$&$2.35\pm0.07$\\
\multirow{2}{*}{$\Xi_c^0 \rightarrow \Sigma^0 K_L$}&$0.97\pm0.16$&$-0.38\pm0.41$&$-0.92\pm0.20$&$-0.06\pm0.43$&$1.40\pm0.36$&$4.59\pm0.92$&$-1.96\pm0.46$\\&$0.97\pm0.16$&$-0.38\pm0.41$&$-0.92\pm0.20$&$-0.06\pm0.43$&$1.40\pm0.36$&$4.59\pm0.92$&$-1.96\pm0.46$\\
\multirow{2}{*}{$\Xi_c^0 \rightarrow \Sigma^{+} \pi^{-}$}&$0.26\pm0.02$&$-0.07\pm0.12$&$-0.96\pm0.03$&$-0.26\pm0.08$&$0.61\pm0.04$&$2.28\pm0.10$&$-1.64\pm0.12$\\&$0.26\pm0.02$&$-0.07\pm0.12$&$-0.96\pm0.03$&$-0.26\pm0.08$&$0.61\pm0.04$&$2.28\pm0.10$&$-1.64\pm0.12$\\
\multirow{2}{*}{$\Xi_c^0 \rightarrow \Sigma^0 \pi^0 $}&$0.34\pm0.03$&$-0.80\pm0.16$&$0.34\pm0.07$&$0.50\pm0.22$&$1.00\pm0.10$&$1.66\pm0.34$&$2.74\pm0.14$\\&$0.34\pm0.03$&$-0.80\pm0.16$&$0.34\pm0.08$&$0.50\pm0.22$&$1.00\pm0.10$&$1.66\pm0.34$&$2.74\pm0.14$\\
\multirow{2}{*}{$\Xi_c^0 \rightarrow \Sigma^0 \eta$}&$0.17\pm0.03$&$-0.99\pm0.07$&$0.09\pm0.18$&$-0.13\pm0.51$&$0.57\pm0.16$&$2.04\pm0.54$&$3.05\pm1.06$\\&$0.17\pm0.03$&$-0.99\pm0.07$&$0.09\pm0.18$&$-0.13\pm0.51$&$0.57\pm0.16$&$2.04\pm0.54$&$3.05\pm0.18$\\
\multirow{2}{*}{$\Xi_c^0 \rightarrow \Sigma^0 \eta'$}&$0.18\pm0.03$&$-0.50\pm0.10$&$0.86\pm0.05$&$-0.11\pm0.24$&$0.72\pm0.12$&$3.37\pm0.43$&$2.10\pm0.12$\\&$0.18\pm0.03$&$-0.50\pm0.10$&$0.86\pm0.05$&$-0.11\pm0.23$&$0.72\pm0.11$&$3.37\pm0.43$&$2.10\pm0.12$\\
\multirow{2}{*}{$\Xi_c^0 \rightarrow \Sigma^- \pi^{+}$}&$1.80\pm0.05$&$-0.73\pm0.03$&$0.68\pm0.03$&$0.02\pm0.04$&$1.89\pm0.05$&$5.30\pm0.13$&$2.39\pm0.04$\\&$1.80\pm0.05$&$-0.73\pm0.03$&$0.68\pm0.03$&$0.02\pm0.04$&$1.89\pm0.05$&$5.30\pm0.13$&$2.39\pm0.04$\\
\multirow{2}{*}{$\Xi_c^0 \rightarrow \Xi^0 K_{S / L}$}&$0.38\pm0.01$&$-0.51\pm0.02$&$0.52\pm0.09$&$-0.69\pm0.07$&$0.51\pm0.06$&$4.26\pm0.12$&$2.35\pm0.10$\\&$0.38\pm0.01$&$-0.51\pm0.02$&$0.52\pm0.09$&$-0.69\pm0.07$&$0.51\pm0.06$&$4.26\pm0.12$&$2.35\pm0.10$\\
\multirow{2}{*}{$\Xi_c^0 \rightarrow p K^{-}$}&$0.31\pm0.02$&$-0.06\pm0.11$&$-0.90\pm0.04$&$-0.44\pm0.07$&$0.61\pm0.04$&$2.28\pm0.10$&$-1.64\pm0.12$\\&$0.31\pm0.02$&$-0.06\pm0.11$&$-0.90\pm0.04$&$-0.44\pm0.07$&$0.61\pm0.04$&$2.28\pm0.10$&$-1.64\pm0.12$\\
\multirow{2}{*}{$\Xi_c^0 \rightarrow n K_{S / L}$}&$0.83\pm0.04$&$-0.36\pm0.01$&$0.37\pm0.07$&$-0.86\pm0.03$&$0.51\pm0.06$&$4.26\pm0.12$&$2.35\pm0.10$\\&$0.83\pm0.04$&$-0.36\pm0.01$&$0.37\pm0.07$&$-0.86\pm0.03$&$0.51\pm0.06$&$4.26\pm0.12$&$2.35\pm0.10$\\
\multirow{2}{*}{$\Xi_c^0 \rightarrow \Lambda^0 \pi^0 $}&$0.07\pm0.01$&$-0.24\pm0.31$&$0.24\pm0.17$&$0.94\pm0.08$&$0.51\pm0.06$&$0.23\pm0.14$&$2.34\pm0.86$\\&$0.07\pm0.01$&$-0.24\pm0.31$&$0.24\pm0.17$&$0.94\pm0.08$&$0.51\pm0.06$&$0.23\pm0.14$&$2.34\pm0.83$\\
\multirow{2}{*}{$\Xi_c^0 \rightarrow \Lambda^0 \eta$}&$0.40\pm0.05$&$-0.47\pm0.22$&$0.78\pm0.12$&$0.41\pm0.12$&$1.10\pm0.10$&$2.05\pm0.22$&$2.11\pm0.26$\\&$0.40\pm0.05$&$-0.47\pm0.22$&$0.78\pm0.12$&$0.41\pm0.12$&$1.10\pm0.10$&$2.05\pm0.22$&$2.11\pm0.26$\\
\multirow{2}{*}{$\Xi_c^0 \rightarrow \Lambda^0 \eta'$}&$0.63\pm0.08$&$-0.62\pm0.07$&$0.64\pm0.13$&$-0.46\pm0.21$&$1.02\pm0.19$&$6.05\pm0.67$&$2.35\pm0.11$\\&$0.63\pm0.08$&$-0.62\pm0.07$&$0.64\pm0.13$&$-0.46\pm0.21$&$1.02\pm0.19$&$6.05\pm0.66$&$2.35\pm0.11$\\
\multirow{2}{*}{$\Xi_c^{+} \rightarrow \Sigma^{+} \pi^0$}&$2.56\pm0.11$&$-0.53\pm0.10$&$0.36\pm0.15$&$-0.77\pm0.12$&$0.62\pm0.16$&$4.94\pm0.20$&$2.54\pm0.16$\\&$2.56\pm0.11$&$-0.53\pm0.10$&$0.36\pm0.15$&$-0.77\pm0.12$&$0.62\pm0.16$&$4.94\pm0.20$&$2.54\pm0.16$\\
\multirow{2}{*}{$\Xi_c^{+} \rightarrow \Sigma^{+} \eta$}&$1.02\pm0.17$&$-0.99\pm0.08$&$0.09\pm0.18$&$-0.13\pm0.51$&$0.81\pm0.22$&$2.89\pm0.76$&$3.05\pm0.18$\\&$1.02\pm0.17$&$-0.99\pm0.08$&$0.09\pm0.18$&$-0.13\pm0.51$&$0.81\pm0.22$&$2.89\pm0.76$&$3.05\pm0.18$\\
\multirow{2}{*}{$\Xi_c^{+} \rightarrow \Sigma^{+} \eta'$}&$1.08\pm0.17$&$-0.50\pm0.10$&$0.86\pm0.05$&$-0.11\pm0.24$&$1.02\pm0.16$&$4.77\pm0.61$&$2.10\pm0.12$\\&$1.08\pm0.17$&$-0.50\pm0.10$&$0.86\pm0.05$&$-0.11\pm0.23$&$1.02\pm0.16$&$4.77\pm0.61$&$2.10\pm0.12$\\

\hline
\hline
\end{tabular}
 }
\end{table*}

\begin{table*}[h!]\footnotesize
\caption{
Same as Table \ref{tab:fitotherScenI} except for yet-observed DCS  modes in Case I.
}
\label{tab:fitother2ScenI}
 \resizebox{\textwidth}{!} 
 {
\centering
\begin{tabular}
{ l | r rrr rr r
}
\hline
\hline
Channel&
$10^{3}\mathcal{B}$~~~~ &$\alpha$~~~~~~~&$\beta$~~~~~~~&$\gamma$~~~~~~
&$|A|$~~~ & $|B|$~~~ & $\delta_P-\delta_S$~~~\\
\hline
\multirow{2}{*}{$\Lambda_c^{+} \rightarrow n K^{+}$}&$0.11\pm0.08$&$-0.93\pm0.27$&$0.38\pm0.66$&$0.02\pm0.40$&$0.14\pm0.06$&$0.35\pm0.17$&$2.75\pm0.71$\\&$0.11\pm0.01$&$-0.93\pm0.05$&$0.38\pm0.12$&$0.02\pm0.14$&$0.14\pm0.01$&$0.35\pm0.04$&$2.75\pm0.12$\\
\multirow{2}{*}{$\Xi_c^0 \rightarrow \Sigma^{-} K^{+}$}&$0.81\pm0.15$&$-0.73\pm0.12$&$0.68\pm0.14$&$0.09\pm0.14$&$0.43\pm0.05$&$1.22\pm0.14$&$2.39\pm0.19$\\&$0.81\pm0.02$&$-0.73\pm0.03$&$0.68\pm0.03$&$0.09\pm0.04$&$0.43\pm0.01$&$1.22\pm0.03$&$2.39\pm0.04$\\
\multirow{2}{*}{$\Xi_c^0 \rightarrow p \pi^{-}$}&$0.19\pm0.03$&$-0.06\pm0.17$&$-0.88\pm0.10$&$-0.47\pm0.17$&$0.14\pm0.02$&$0.52\pm0.06$&$-1.64\pm0.20$\\&$0.19\pm0.01$&$-0.06\pm0.11$&$-0.88\pm0.04$&$-0.47\pm0.07$&$0.14\pm0.01$&$0.52\pm0.02$&$-1.64\pm0.12$\\
\multirow{2}{*}{$\Xi_c^0 \rightarrow n \pi^0 $}&$0.09\pm0.02$&$-0.06\pm0.17$&$-0.88\pm0.10$&$-0.47\pm0.17$&$0.10\pm0.02$&$0.37\pm0.04$&$-1.64\pm0.20$\\&$0.09\pm0.01$&$-0.06\pm0.11$&$-0.88\pm0.04$&$-0.47\pm0.07$&$0.10\pm0.01$&$0.37\pm0.02$&$-1.64\pm0.12$\\
\multirow{2}{*}{$\Xi_c^0 \rightarrow n \eta$}&$0.56\pm0.06$&$-0.46\pm0.08$&$0.62\pm0.08$&$-0.64\pm0.04$&$0.21\pm0.02$&$1.06\pm0.04$&$2.21\pm0.14$\\&$0.56\pm0.05$&$-0.46\pm0.06$&$0.62\pm0.06$&$-0.64\pm0.06$&$0.21\pm0.02$&$1.06\pm0.05$&$2.21\pm0.10$\\
\multirow{2}{*}{$\Xi_c^0 \rightarrow n \eta'$}&$0.31\pm0.09$&$-0.46\pm0.22$&$0.64\pm0.14$&$-0.61\pm0.13$&$0.19\pm0.04$&$1.04\pm0.17$&$2.20\pm0.30$\\&$0.31\pm0.05$&$-0.46\pm0.07$&$0.64\pm0.15$&$-0.61\pm0.17$&$0.19\pm0.04$&$1.04\pm0.13$&$2.20\pm0.12$\\
\multirow{2}{*}{$\Xi_c^{+} \rightarrow n \pi^{+}$}&$0.56\pm0.09$&$-0.06\pm0.17$&$-0.88\pm0.10$&$-0.47\pm0.17$&$0.14\pm0.02$&$0.52\pm0.06$&$-1.64\pm0.20$\\&$0.56\pm0.04$&$-0.06\pm0.11$&$-0.88\pm0.04$&$-0.47\pm0.07$&$0.14\pm0.01$&$0.52\pm0.02$&$-1.64\pm0.12$\\
\multirow{2}{*}{$\Xi_c^{+} \rightarrow \Sigma^0 K^{+}$}&$1.20\pm0.22$&$-0.73\pm0.12$&$0.68\pm0.14$&$0.09\pm0.14$&$0.31\pm0.04$&$0.86\pm0.10$&$2.39\pm0.19$\\&$1.20\pm0.03$&$-0.73\pm0.03$&$0.68\pm0.03$&$0.09\pm0.04$&$0.31\pm0.01$&$0.86\pm0.02$&$2.39\pm0.04$\\
\multirow{2}{*}{$\Xi_c^{+} \rightarrow p \pi^0$}&$0.28\pm0.04$&$-0.06\pm0.17$&$-0.88\pm0.10$&$-0.47\pm0.17$&$0.10\pm0.02$&$0.37\pm0.04$&$-1.64\pm0.20$\\&$0.28\pm0.02$&$-0.06\pm0.11$&$-0.88\pm0.04$&$-0.47\pm0.07$&$0.10\pm0.01$&$0.37\pm0.02$&$-1.64\pm0.12$\\
\multirow{2}{*}{$\Xi_c^{+} \rightarrow p \eta$}&$1.68\pm0.17$&$-0.46\pm0.08$&$0.62\pm0.08$&$-0.64\pm0.04$&$0.21\pm0.02$&$1.06\pm0.04$&$2.21\pm0.14$\\&$1.68\pm0.13$&$-0.46\pm0.06$&$0.62\pm0.06$&$-0.64\pm0.06$&$0.21\pm0.02$&$1.06\pm0.05$&$2.21\pm0.10$\\
\multirow{2}{*}{$\Xi_c^{+} \rightarrow p \eta'$}&$0.92\pm0.27$&$-0.46\pm0.22$&$0.64\pm0.14$&$-0.61\pm0.13$&$0.19\pm0.04$&$1.04\pm0.17$&$2.20\pm0.30$\\&$0.92\pm0.16$&$-0.46\pm0.07$&$0.64\pm0.15$&$-0.61\pm0.17$&$0.19\pm0.04$&$1.04\pm0.13$&$2.20\pm0.12$\\

\multirow{2}{*}{$\Xi_c^{+} \rightarrow \Lambda^0 K^{+}$}&$0.42\pm0.09$&$-0.13\pm0.15$&$-0.43\pm0.19$&$-0.89\pm0.08$&$0.06\pm0.02$&$0.67\pm0.08$&$-1.87\pm0.40$\\&$0.42\pm0.04$&$-0.13\pm0.11$&$-0.43\pm0.10$&$-0.89\pm0.05$&$0.06\pm0.01$&$0.67\pm0.03$&$-1.87\pm0.25$\\
\hline
\hline
\end{tabular}
 }
\end{table*}

\end{document}